\documentclass{article}
\usepackage{graphicx}

\usepackage{amsmath}
\usepackage{amssymb}
\usepackage{mathtools}
\usepackage{amsthm}
\usepackage{multirow}
\usepackage{makecell}

\usepackage[preprint]{neurips_2025}

\usepackage[utf8]{inputenc} % allow utf-8 input
\usepackage[T1]{fontenc}    % use 8-bit T1 fonts
\usepackage{hyperref}       % hyperlinks
\usepackage{url}            % simple URL typesetting
\usepackage{booktabs}       % professional-quality tables
\usepackage{amsfonts}       % blackboard math symbols
\usepackage{nicefrac}       % compact symbols for 1/2, etc.
\usepackage{microtype}      % microtypography
\usepackage{xcolor}         % colors

\title{Masked Language Models are Good Heterogeneous Graph Generalizers}

\author{Jinyu Yang$^{1}$, Cheng Yang$^{1}$, Shanyuan Cui$^{1}$\\\textbf{Zeyuan Guo$^{1}$, Liangwei Yang$^2$, Muhan Zhang$^3$, Zhiqiang Zhang$^4$, Chuan Shi$^{1}$}\\
\\
$^1$ School of Computer Science, Beijing University of Posts and Telecommunications,\\
$^2$ Department of Computer Science, University of Illinois Chicago,\\
$^3$ Institute for Artificial Intelligence, Peking University\\
$^4$ Ant Group, Beijing, China
\\
\texttt{jinyu.yang@bupt.edu.cn},\quad \texttt{yangcheng@bupt.edu.cn}, \quad
\texttt{cuishanyuanai@bupt.edu.cn} \\
}

\begin{document}

\maketitle

\begin{abstract}
  Heterogeneous graph neural networks (HGNNs) excel at capturing structural and semantic information in heterogeneous graphs (HGs), while struggling to generalize across domains and tasks. With the rapid advancement of large language models (LLMs), a recent study explored the integration of HGNNs with LLMs for generalizable heterogeneous graph learning. However, this approach typically encodes structural information as HG tokens using HGNNs, and disparities in embedding spaces between HGNNs and LLMs have been shown to bias the LLM’s comprehension of HGs. Moreover, since these HG tokens are often derived from node-level tasks, the model’s ability to generalize across tasks remains limited. To this end, we propose a simple yet effective Masked Language Modeling-based method, called MLM4HG. MLM4HG introduces metapath-based textual sequences instead of HG tokens to extract structural and semantic information inherent in HGs, and designs customized textual templates to unify different graph tasks into a coherent cloze-style \texttt{<mask>} token prediction paradigm. Specifically, MLM4HG first converts HGs from various domains to texts based on metapaths, and subsequently combines them with the unified task texts to form a HG-based corpus. Moreover, the corpus is fed into a pretrained LM for fine-tuning with a constrained target vocabulary, enabling the fine-tuned LM to generalize to unseen target HGs. Extensive cross-domain and multi-task experiments on four real-world datasets demonstrate the superior generalization performance of MLM4HG over state-of-the-art methods in both few-shot and zero-shot scenarios. Our code is available at \href{https://github.com/BUPT-GAMMA/MLM4HG}{https://github.com/BUPT-GAMMA/MLM4HG}.

\end{abstract}

\section{Introduction}
Heterogeneous graphs (HGs) containing diverse node and edge types are ubiquitous in real-world scenarios, such as social networks \cite{jinyu:1}, knowledge graphs \cite{jinyu:2}, recommender systems \cite{jinyu:3}, and biological networks \cite{jinyu:4}. %Considering the structural and semantic information inherent in HGs \cite{jinyu:6, jinyu:5}, metapaths have been widely used in lots of heterogeneous graph neural networks (HGNNs). Moreover, several self-supervised learning (SSL) methods \cite{jinyu:48, jinyu:45} have also been proposed to achieve heterogeneous graph representation learning. 
Embedding the structure and attribute information of HGs into a low-dimensional vector space, numerous heterogeneous graph neural networks (HGNNs) \cite{jinyu:6, jinyu:5} and self-supervised learning (SSL) \cite{jinyu:48, jinyu:45} methods have been proposed and demonstrated promising performance in plenty of graph tasks. Nevertheless, these methods are primarily designed for a single HG or a specific graph task, significantly limiting their generalization across different domains and tasks \cite{jinyu:10}. Consequently, the development of generalizable heterogeneous graph learning methods has emerged as a central focus \cite{jinyu:41}.

With the great success of large language models (LLMs), researchers turn to harnessing LLMs for generalizable graph learning \cite{jinyu:11}. To leverage the powerful understanding ability of LLMs and preserve the graph structural information, current researches either utilize GNNs to encode graphs into graph tokens \cite{jinyu:11} or converts graphs into texts \cite{jinyu:19}, after which the tokens or texts are fed into LLMs for inference. These approaches are mainly tailored to homogeneous graphs and overlook the rich structural and semantic information inherent in HGs. A recent research, HiGPT~\cite{jinyu:10}, introduces heterogeneity-aware graph instructions to help LLM deal with HGs. However, the structural information is extracted via an HGNN and encoded as HG tokens. The inherent mismatch between the embedding spaces of HGNNs and LLMs \cite{jinyu:13} introduces biases into the LLM's understanding of HGs. Moreover, HiGPT primarily focuses on node-level tasks, overlooking other essential graph tasks such as link prediction, which hinders its multi-task generalization ability and leads to suboptimal performance.

% 第一个问题：嵌入空间不适配 -> 表达形式统一
% 第二个问题：无法实现多任务学习 -> 任务统一
% To address these problems, we aim to eliminate the reliance on HGNNs, and rely solely on language models to characterize metapath-based sequential information. To this end, we propose a simple yet effective Masked Language Modeling-based method for generalizable Heterogeneous Graph learning, namely MLM4HG. Firstly, we use a metapath-based graph-to-text method to extract intrinsic structural and semantic properties of HGs, and convert HGs from various domains into texts. Secondly, to enable our model to generalize across different graph tasks, we customize textual templates to unify them into task sequences. These textual sequences are then combined to form an HG-based text corpus. Thirdly, after replacing the label tokens (\textit{e.g.,} ground truth node categories) with \texttt{<mask>} in the corpus, a pretrained masked language model is fine-tuned to predict the masked tokens using a constrained target vocabulary. The fine-tuned LM can subsequently be adapted to unseen target HGs across different graph tasks.
To address these problems, we propose a simple yet effective Masked Language Modeling-based method for generalizable Heterogeneous Graph learning, namely MLM4HG. To enhance cross-domain generalization, our method forgoes HGNNs and instead extracts the structural and semantic information of HGs from metapath-based sequences. Furthermore, to support multi-task generalization, we reformulate various graph tasks into a unified cloze-style \texttt{<mask>} prediction paradigm using customized textual templates.
%Different graph tasks are also cast as a uniform cloze-style \texttt{<mask>} token prediction paradigm using customized textual templates. Relying solely on language models as the backbone, we enable generalizable heterogeneous graph learning. 
Specifically, we first employ a metapath-based graph-to-text method to extract the structural and semantic properties of HGs and convert them into texts. Next, we customize textual templates to reformulate different tasks into task sequences. These textual sequences are then combined to form an HG-based text corpus. Finally, after replacing the label tokens (\textit{e.g.,} ground truth node categories) with \texttt{<mask>} in the corpus, a pretrained masked language model is fine-tuned to predict the masked tokens using a constrained target vocabulary. The fine-tuned LM can subsequently be adapted to unseen target HGs across different graph tasks.

We conduct extensive cross-domain and multi-task experiments by fine-tuning the model on source HGs across different graph tasks and adapting it to an unseen target HG. The evaluation is performed on four real-world datasets from distinct domains and graph tasks in both few-shot and zero-shot scenarios. The results demonstrate that MLM4HG outperforms the best-performing baselines by an average of 18.08\% in Micro-F1 and 7.10\% in Macro-F1 in few-shot settings, as well as 15.22\% in Micro-F1 and 8.04\% in Macro-F1 in zero-shot settings. Our contributions are summarized as follows:

$\bullet$ To the best of our knowledge, we are the first to standardize HG representations and reframe different graph tasks as a unified cloze-style \texttt{<mask>} token prediction paradigm, establishing a generalizable learning framework fully powered by a masked language model.

$\bullet$ We propose MLM4HG, which first transforms HGs from various domains into texts based on metapaths, and subsequently combines them with the unified task texts to form a HG-based corpus. Then, the corpus is used to fine-tune a pretrained masked language model with a constrained target vocabulary to enable generalization.

$\bullet$ We conduct extensive cross-domain and multi-task experiments on four real-world datasets, demonstrating MLM4HG’s superior generalization performance across domains and tasks.

\section{Related Work}
\subsection{Heterogeneous Graph Representation Learning}
Heterogeneous graph neural networks (HGNNs)~\cite{jinyu:30, jinyu:51} and self-supervised learning (SSL)~\cite{jinyu:52} are effective methods for heterogeneous graph representation learning. 

\noindent\textbf{Existing HGNNs} can be broadly categorized into two types. Relation-based methods \cite{jinyu:33, jinyu:31, jinyu:32} capture heterogeneous structural information by treating different edge types as distinct relations. For example, RHINE \cite{jinyu:31} distinguish the various relations into Affiliation Relations (ARs) with one-centered by-another structures and Interaction Relations (IRs) with peer-to-peer structures. Considering the semantic information inherent in HGs, lots of metapath-based methods \cite{jinyu:34, jinyu:6, jinyu:5} have emerged in recent years. For instance, HAN \cite{jinyu:5} adopts hierarchical attention to aggregate information at both the node and semantic levels. 

\noindent\textbf{SSL methods} include contrastive methods and generative methods. Contrastive methods~\cite{jinyu:47, jinyu:45} aim to maximize mutual information between different views. For example, HeCo \cite{jinyu:45} employs network schema and meta-path views to assist the model in learning high-level node embeddings. Generative methods~\cite{jinyu:46, jinyu:48}, on the other hand, leverage graph reconstruction and property prediction to extract general structural and attribute semantics. For instance, GPT-GNN \cite{jinyu:46} utilizes self-supervised attribute generation and edge generation tasks to capture the inherent dependencies between node attributes and graph structure. However, these methods focus on a single HG domain or a specific task, lacking the capacity to generalize across diverse HGs and tasks.

\subsection{Large Language Models for Graphs}
With the rapid advancements of large language models (LLMs) \cite{jinyu:36, jinyu:27}, researchers are increasingly focusing on LLM-enhanced graph learning \cite{jinyu:37}. Recent approaches can be categorized into three main types. The first category utilizes GNNs to generate graph tokens, which aid LLMs in capturing structural information \cite{jinyu:38, jinyu:39, jinyu:11}. The second category leverages LLMs to generate labels, features, and prompts for subsequent GNN training \cite{jinyu:43, jinyu:41, jinyu:42}. The third category attempts to use LLMs instead of GNNs as the backbone \cite{jinyu:12, jinyu:13, jinyu:40}, aiming to directly convert graphs into LLM-compatible sequences \cite{jinyu:19} and utilize various instructions to assist LLMs in conducting downstream graph tasks. However, these approaches focus on homogeneous graphs and overlook the heterogeneous graphs prevalent in real-world scenarios. Although the recent HiGPT \cite{jinyu:10} makes an initial attempt at recognizing heterogeneity, its reliance on HGNNs introduces biases in the LLM's understanding of HGs. Additionally, its focus on node-level tasks limits its ability to support multi-task learning, which further leads to suboptimal performance.

\section{Preliminaries}
This section first introduces basic concepts and then presents the learning scenarios to evaluate the generalization ability.

\noindent\textit{Definition 1.} \textbf{Heterogeneous Graph}. A heterogeneous graph is defined as $\mathcal{G} = (\mathcal{V}, \mathcal{E}, \mathcal{X})$. $\mathcal{V}$ represents the set of nodes with a node type mapping function $\varphi : \mathcal{V} \mapsto \mathcal{A}$, and $\mathcal{E}$ denotes the set of edges with an edge type mapping function $\psi : \mathcal{E} \mapsto \mathcal{R}$. $\mathcal{A}$ and $\mathcal{R}$ respectively denote the sets of node and edge types, where $|\mathcal{A}| + |\mathcal{R}| > 2$. $\mathcal{X}$ represents a textual attribute set of nodes, and each node $v_i \in \mathcal{V}$ is associated with a textual attribute feature $x_i \in \mathcal{X}$.

\noindent\textit{Definition 2.} \textbf{Metapath}. A metapath $P$ is defined as a path template in the form of $A_1 \stackrel{R_1}{\longrightarrow} A_2 \stackrel{R_2}{\longrightarrow} \cdots \stackrel{R_l}{\longrightarrow} A_{l+1}$, which describes a composite relation $R = R_1 \circ R_2 \circ \cdots \circ R_l$ between node types $A_1$ and $A_{l+1}$, where $\circ$ denotes the composition operator on relations.

\noindent\textbf{Learning Scenarios}. To evaluate the model's generalization ability across various domains and tasks, following previous works \cite{jinyu:12, jinyu:10}, we conduct multi-task training on various source HGs and evaluate it on a target HG. The source and target HGs are from distinct domains. We refer to this configuration as the \textit{cross-domain and multi-task} setting. In the target HG, only a few or even no nodes are labeled, which can be abstracted as few-shot or zero-shot scenarios. The few-shot scenario is often described as an \textit{N}-way \textit{K}-shot setting: \textit{N}-way refers to the number of classes, and \textit{K}-shot indicates that there are \textit{K} labeled nodes per class, while in the zero-shot case, \textit{K} is set to 0. Our goal is to extract generalizable knowledge under the \textit{cross-domain and multi-task} setting, and subsequently transfer it to an unseen target HG across different graph tasks in both few-shot and zero-shot scenarios.

\begin{figure}[t]
    \centering
    \includegraphics[width=1\linewidth]{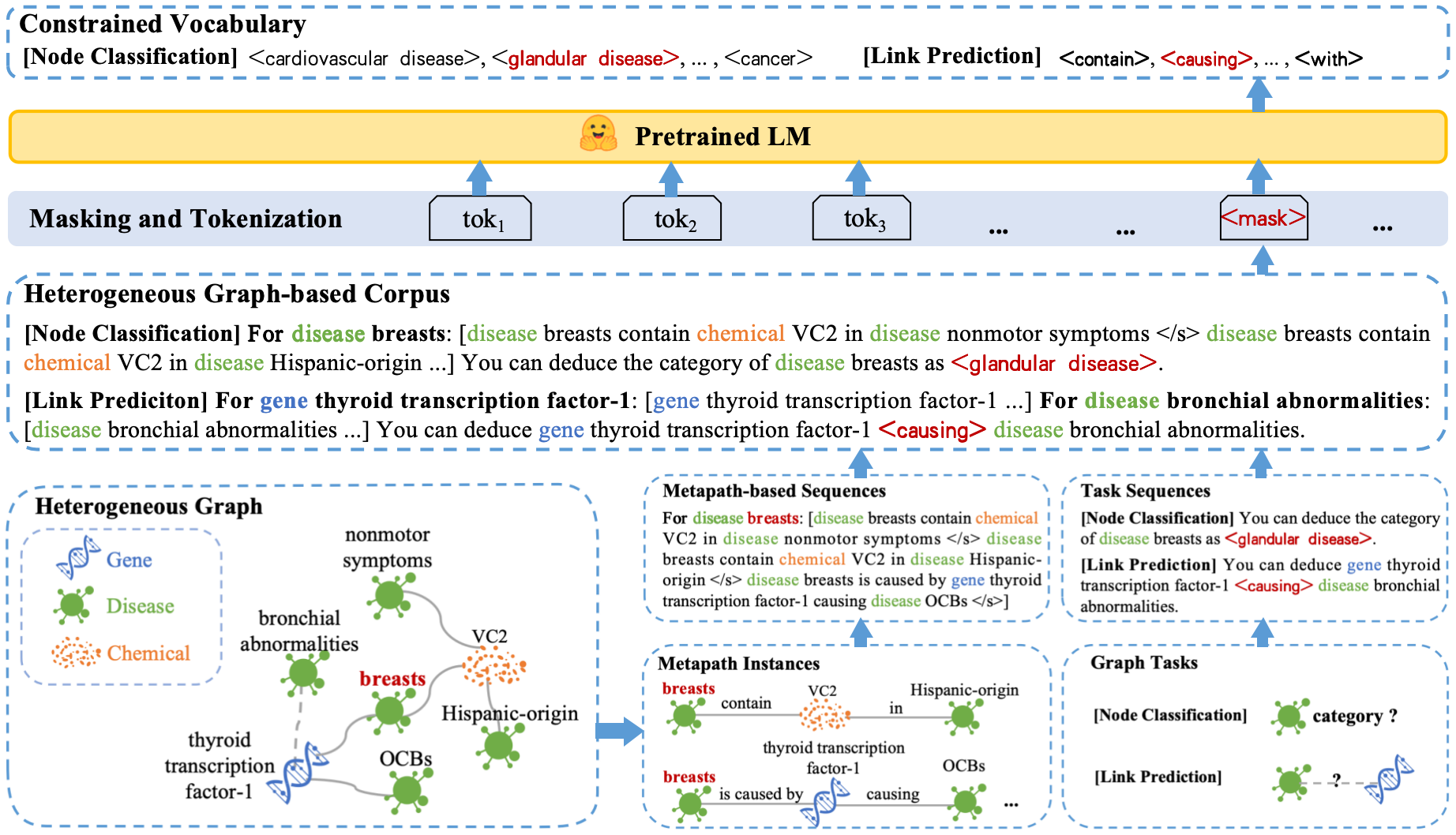}
    \caption{The overall framework of the proposed MLM4HG. We first convert HGs from different domains as well as the corresponding graph tasks into a unified textual format, forming a HG-based text corpus. Then we fine-tune a pretrained LM on the corpus by predicting masked label tokens. Finally, the LM can be adapted to unseen target HGs for generalization. Note that our performance does not solely rely on the prior knowledge lying in pretrained LMs. In Section \ref{sec:explore}, we anonymize nodes in the corpus by their ids, and can still achieve satisfying results.}
    \label{fig:framework}
\end{figure}

\section{Methodology}
\label{main:methodology}
This section presents our proposed MLM4HG, with its framework shown in Figure \ref{fig:framework}. 
% Inspired by the masked language modeling (MLM) \cite{jinyu:16}, we utilize a masked language model as the backbone and apply an enhanced MLM for generalizable heterogeneous graph learning. First, we design a masked heterogeneous graph corpus construction method to unify the representation across various HGs and graph tasks, facilitating the model's cross-domain training and seamless task switching. These corpora will then be used as inputs to the masked language model. Furthermore, during the training stage, we design a graph task-related constrained vocabulary for the prediction of \texttt{<mask>} tokens, which enhances the LM's ability to effectively capture the intricate relationships between the training targets (\textit{e.g.,}, node labels in node classification tasks or link categories in link prediction tasks) and the masked graph corpus. A further time complexity analysis of MLM4HG is provided in Appendix \ref{app:time complexity}.
We first convert HGs from various domains into texts based on metapaths, and unify different graph tasks by textual templates. The above texts are then combined as a heterogeneous graph-based corpus. Afterward, we fine-tune a pretrained LM on the corpus by predicting the masked label tokens. Finally, the LM can be adapted to unseen target HGs for generalizable HG learning. %construction method to unify the textual representation across domains/tasks and an enhanced MLM training method for 

\subsection{Heterogeneous Graph-based Corpus Construction}
To help LMs generalize across different domains and tasks, we construct a heterogeneous graph-based corpus by converting relevant information into text sequences. Specifically, the HG-based corpus contains entries from different HGs and tasks, and each entry is composed of a task sequence and relevant nodes' metapath-based sequences.

\noindent\textbf{Metapath-based Sequences.} Heterogeneous graphs contain rich structural and semantic information, which can be extracted through metapaths~\cite{jinyu:6, jinyu:5}. For each node in a heterogeneous graph, metapath instances can be sampled based on its neighborhood, and subsequently textualized as metapath-based sequences. In this way, our method unifies the formats of different HGs, and enables the generalization ability across domains.

Specifically, we first initialize metapaths for each heterogeneous graph based on node and edge types. For each node $v$, we will consider all the metapaths starting with node type $\varphi(v)$, and then sample metapath instances  in the form of $v \stackrel{e_1}{\longrightarrow} v_1 \stackrel{e_2}{\longrightarrow} \cdots \stackrel{e_l}{\longrightarrow} v_{l}$. Then each node $v_i$ in the metapath instance is textualized by concatenating its node type $\varphi(v_i)$ and textual attribute $x_i$ as $T(v_i)$, such as $<$ \texttt{disease breasts} $>$. Further, each metapath instance is textualized in the form $[T(v), \psi(e_1), T(v_1), ... , T(v_{l-1}), \psi(e_l), T(v_{l})]$, where $\psi(e)$ is the edge type of $e$. Finally, all metapath instances associated with node $v$ are concatenated with special token \texttt{</s>} for separation, denoted as text sequence $MS_v$. For example, the metapath-based sequence for disease node \texttt{breasts} is illustrated in Figure \ref{fig:framework}.
%}
%While edge $e_j$ is represented by edge type $\psi(e_j)$ as $T(e_j)$, such as $<$ \texttt{contain} $>$ $j \in [1, l]$

\noindent\textbf{Task Sequences.} To enable the LM to generalize across different graph tasks, we unify task formats by the following textual templates:% Specifically, we replace the training targets with \texttt{<mask>} tokens for different tasks, as illustrated in Figure \ref{fig:framework} (a).

In the classification task, the target is to predict the category of an unlabeled object. Therefore, we standardize the classification task format as $<$ \texttt{You can deduce the category of [object] as <target>.} $>$, where \texttt{[object]} refers to the textualized object to be classified and \texttt{<target>} represents the category label. For example, for the classification of disease node \texttt{breasts}, the corresponding task sequence is formatted as $<$ \texttt{You can deduce the category of disease breasts as <glandular disease>.} $>$.%Specifically, in node classification task, \texttt{[object]} corresponds to the textualization $T(v)$ of node $v$. 

In the link prediction task, the target shifts to determining whether an edge exists between two nodes and identifying its type. Therefore, we standardize the link prediction task format as $<$ \texttt{You can deduce [src] \texttt{<target>} [dst].} $>$, where \texttt{[src]} and \texttt{[dst]} respectively represent the textualization $T(v_1)$ and $T(v_2)$ of source node $v_1$ and destination node $v_2$. \texttt{<target>} denotes the edge type between $v_1$ and $v_2$. For example, for the link prediction between the source node gene \texttt{thyroid transcription factor-1} and the destination node disease \texttt{bronchial abnormalities}, the corresponding task sequence is formatted as $<$ \texttt{You can deduce gene thyroid transcription factor-1 \texttt{<causing>} disease bronchial abnormalities.} $>$. Utilizing these templates, we can automatically collect task sequences $TS$ covering all tasks and their examples.

\noindent\textbf{Corpus Organization.} Each entry in our HG-based corpus $\mathcal{I}$ consists of a specific task sequence with the metapath-based sequences of relevant nodes as prefix. Specifically, the $j$-th example of the $i$-th task is represented as entry $\mathcal{I}^i_j = [MS_{v_1}, MS_{v_2}, ..., MS_{v_M}, TS^i_j]$, where $M$ is the number of related nodes. For example, $M = 1$ for the node classification task, and $M = 2$ for link prediction.
%graph classification task, $n$ is the number of nodes in the HG.

\subsection{Masked Language Modeling for Generalizable HG Learning}
Given the HG-based corpus, we fine-tune a pretrained LM by multiple graph tasks via masked token prediction. Then the LM can generalize to tasks on unseen target HGs with limited annotations.

\noindent\textbf{Masking and Tokenization.} Different from the random masking strategy in BERT \cite{jinyu:16}, we only replace \texttt{<target>} tokens with \texttt{<mask>} for each entry. For example, in the node classification task for disease \texttt{breasts}, the target category \texttt{<glandular disease>} will be replaced with \texttt{<mask>}. %Then, the masked entry is tokenized as $Toks = \left\{tok_1, \ldots, tok_{m-1}, \texttt{<mask>}, tok_{m+1}, \ldots, tok_L\right\}$.

\noindent\textbf{MLM with Constrained Vocabulary.} The targets of graph tasks typically focus on a constrained set. For instance, in the node classification task of disease \texttt{breasts} shown in Figure \ref{fig:framework}, we aim to determine whether the \texttt{<mask>} token corresponds to a disease category, such as \texttt{<cardiovascular disease>}, \texttt{<glandular disease>}, or \texttt{<cancer>}, rather than considering the entire vocabulary as in typical language model tasks. Therefore, we adopt a constrained output vocabulary for each graph task. Specifically, each entry $\mathcal{I}^i_j$ is paired with a constrained target vocabulary $\mathcal{C}^i_j$, facilitating the LM to generalize across different datasets and tasks. This design choice closely mirrors the use of instruction-based output constraints in LLMs \cite{jinyu:11, jinyu:10}. For example, in the IMDB node classification task, the LLM is guided by the instruction: "\texttt{Which of the following classes does this movie belong to: action, comedy, drama?}"

We denote the masked token sequence of entry $\mathcal{I}^i_j$ as $\text{Tok}_j^i$, and the corresponding ground truth target as $\text{Tar}_j^i\in \mathcal{C}^i_j$. The optimization objective is defined by the classical cross entropy loss \cite{jinyu:16, jinyu:17} as:
\begin{equation}
\mathcal{L}_\text{MLM}=-\frac{1}{|\mathcal{I}|} \sum_{i,j}  \log p_\text{MLM}\left(\text{Tar}_j^i \mid \text{Tok}_j^i\right),
\label{eq:eq_1}
\end{equation}
where $p_\text{MLM}$ is the language modeling probability within the constrained vocabulary $\mathcal{C}^i_j$.

\noindent\textbf{Generalization to Unseen Target HGs.} To adapt the fine-tuned LM to unseen target HGs, we first transform the target HGs into the same format as the corpus. Then for zero-shot scenarios, we directly ask the LM to predict within the corresponding constrained vocabulary; for few-shot scenarios, we will update the parameters of LM in the same way as our fine-tuning process before inference. Detailed analyses of the scalability of MLM4HG are presented in Appendix~\ref{app:time complexity}.

\subsection{Discussion}
\label{sec:explore}
\noindent\textbf{Why Masked LMs Instead of LLMs?} MLM4HG reformulates various tasks into a unified cloze-style \texttt{<mask>} token prediction paradigm, which essentially corresponds to classification. Prior work~\cite{jinyu:54} shows that, compared to causal language modeling-based LLMs, masked LMs are better suited for such tasks both theoretically and empirically. Consequently, masked LMs align more naturally with the cloze-style formulation. Empirically, they achieve superior performance (Sections~\ref{overall performance} and~\ref{zero-shot performance}) while incurring significantly lower computational overhead (Appendix~\ref{app:Impact of LMs}).

\noindent\textbf{Relying Solely on Pretrained LM's Prior Knowledge?} 
As discussed in this subsection, we adopt a pretrained language model with inherent prior knowledge as the backbone. To access whether the LM can effectively capture structural information embedded in metapath-based sequences, which is independent of its original knowledge, and generalize to the target dataset, we conduct an exploratory experiment on the supervised node classification task using the IMDB dataset. 

\begin{figure}[t]
    \centering
    \includegraphics[width=1\linewidth]{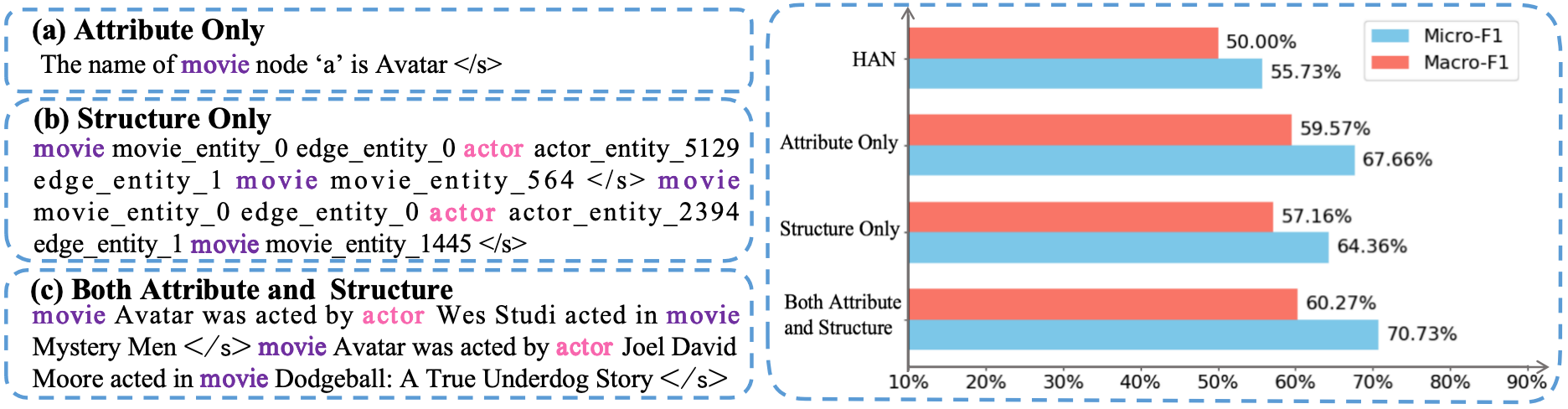}
    \caption{Metapath-based sentences in (a) Attribute Only, (b) Structure Only and (c) Both Attribute and Structure settings. Experiments are conducted on the IMDB node classification task.}
    \label{fig:structure_and_attribute}
\end{figure}

As shown in Figure \ref{fig:structure_and_attribute}, we process the metapath-based sequences into three types: \textbf{Attribute Only}, \textbf{Structure Only}, and \textbf{Both Attribute and Structure}. In \textbf{Attribute Only}, only attribute information is included. In \textbf{Structure Only}, all nodes are anonymized with no attribute information, while in \textbf{Both Attribute and Structure}, both structure and attribute information are provided. The results are shown in Figure \ref{fig:structure_and_attribute}. In the 20\% supervised learning scenarios, all three input types outperform the metapath-based HGNN baseline HAN \cite{jinyu:5}, with \textbf{Both Attribute and Structure} achieving the best results. This highlights the effectiveness of metapath-based sequences in allowing the LM to integrate both structural and semantic information from various source heterogeneous graphs. While the LM’s prior knowledge contributes to the strong performance of \textbf{Attribute Only}, the competitive results of \textbf{Structure Only} indicate that our approach does not merely rely on pretrained knowledge. Instead, the LM is able to extract and leverage structural signals from metapath-based sequences for downstream tasks.

\section{Experiments and Analysis}
To validate the effectiveness of our MLM4HG model, we conduct extensive cross-domain and multi-task experiments on four real-world heterogeneous graph datasets from various domains under diverse settings to answer the following research questions:

\textbf{RQ1:} How well does MLM4HG \textit{\textbf{generalize across domains}} in few-shot scenarios?

\textbf{RQ2:} How well does MLM4HG \textit{\textbf{generalize across tasks}} in zero-shot scenarios?

\textbf{RQ3:} What is \textit{\textbf{the impact of model components and learning scenarios}} on performance?

% \textbf{RQ4:} What is \textit{\textbf{the impact of adopting different LMs or LLMs as backbones}}?

\textbf{RQ4:} How effective are metapath-based sequences \textit{\textbf{compared to other graph-to-text methods}}?

Additionally,  we further investigate the impact of using different LMs or LLMs as backbones in Appendix \ref{app:Impact of LMs}. We also analyze the impact of different hyperparameters (Appendix~\ref{app:hyper_para}) and provide a case study (Appendix~\ref{app:case study}) with a visual analysis of attention scores from the final layer of MLM4HG across various graph tasks.

\subsection{Experimental Settings}
\label{main:exp setting}
\noindent\textbf{Datasets.} We conduct extensive experiments on four real-world heterogeneous graph benchmark datasets from different domains: IMDB, DBLP \cite{jinyu:5}, YELP, and PubMed \cite{jinyu:9}. More details can be found in Appendix \ref{app:dataset}.

\noindent\textbf{Baselines.} To comprehensively evaluate the performance of our MLM4HG, we compare it with the following five representative and state-of-the-art methods from three different categories. The first category is represented by the HGNN method HAN \cite{jinyu:5}. The second category is the heterogeneous graph self-supervised learning method HeCo \cite{jinyu:45}. The third category includes LM and LLM-enhanced graph methods, including WalkLM \cite{jinyu:19}, GraphGPT \cite{jinyu:11}, and HiGPT \cite{jinyu:10}, in which HiGPT is specifically designed for heterogeneous graphs, while the remaining methods are tailored for homogeneous graphs.

To prevent potential concerns regarding fairness due to our use of constrained vocabularies, we provide additional clarification here. HiGPT and GraphGPT restrict the output space of LLMs through instructions such as: "\texttt{Which class does this movie belong to: action, comedy, drama?}", a strategy that successfully bounds over 99\% of the generated outputs. Similarly, other baselines adopt task-specific prediction heads that inherently limit the output dimensions. These practices collectively ensure a fair basis for comparison. Further details are provided in Appendix~\ref{app:baselines}.

\noindent\textbf{Evaluation Protocols.} We adopt the \textit{cross-domain and multi-task} setting during fine-tuning. For instance, when IMDB serves as the target HG, we fine-tune the pretrained language model on DBLP, PubMed, and YELP as source HGs, using both node classification and link prediction tasks. The model's generalization ability is then evaluated on the unseen target HG across these two tasks. For the node classification task, we evaluate on both few-shot and zero-shot scenarios. The few-shot setting follows an \textit{N}-way \textit{K}-shot protocol with \textit{K} in \{1, 5, 20, 40\} \cite{jinyu:10}. For the link prediction task, we assess model performance under the zero-shot scenario. Since we unify different graph tasks into the \texttt{<mask>} token prediction paradigm, the link prediction task in our experiments is essentially treated as an edge classification task. Accordingly, both node classification and link prediction are evaluated using Micro-F1 and Macro-F1 scores. (We also report AUC and AP metrics for the link prediction task in Appendix~\ref{auc and ap for lp}.) All reported results are averaged over five independent runs.

\noindent\textbf{Implementation Details.} The LM used in our experiments is pretrained DistilRoBERTa \cite{jinyu:17}. To automatically generate diverse metapaths while reducing reliance on manual design, we comprehensively traverse all edge types to construct candidate metapaths, as shown in Appendix \ref{app:dataset}. For each node in HGs, we randomly sample 3 instances per metapath. This design choice is motivated by our observation (Appendix \ref{app:hyper_para}) that longer metapath-based sequences enable the LM to better capture intricate structural and semantic information. Notably, sampling three instances yields sequence lengths that closely align with the model's maximum input capacity of 512 tokens, thereby fully utilizing its contextual modeling capabilities. The LM is fine-tuned for three epochs, and the other hyperparameters of MLM4HG are optimized through grid-searching for best performance. The baseline parameters are initially set according to the values reported in the original papers and then optimized through grid-searching to achieve optimal performance. All hyperparameter settings are provided in Appendix \ref{app:implement details}, and a detailed hyperparameter study is available in Appendix \ref{app:hyper_para}.

\subsection{Cross-domain Generalization Analysis (RQ1)}
\label{overall performance}
To validate the model's generalization ability on datasets from distinct domains, we fine-tune the model under the \textit{cross-domain and multi-task} setting and then perform node classification experiments on target HGs in few-shot scenarios. 

The overall experimental results are presented in Table \ref{tab:main cd-ct few-shot}. (1) MLM4HG consistently outperforms a range of state-of-the-art baselines across 16 evaluation groups on four datasets, achieving an average gain of 18.08\% in Micro-F1 and 7.10\% in Macro-F1. (2) Notably, MLM4HG with only \textit{1}-shot supervision surpasses other methods with \textit{40}-shot settings on the IMDB and PubMed datasets, indicating its strong ability to capture transferable knowledge from source HGs and effectively generalize to target HGs with minimal supervision. (3) HiGPT achieves the second-best performance in most cases and outperforms GraphGPT, benefitting from its heterogeneous graph instruction tuning, which helps it recognize relation types more effectively. However, the reliance on HG tokens generated by HGNNs introduces embedding space mismatch, and the lack of metapath-level semantics further hinders its performance. (4) WalkLM performs better than HiGPT on PubMed, likely because its random walk-based sequences enable the LM to more directly capture graph structures, which is especially useful for PubMed’s complex biomedical data. However, WalkLM fails to account for generalization and omits metapaths, which ultimately leads to suboptimal performance.

\subsection{Multi-task Generalization Analysis (RQ2)}
\label{zero-shot performance}
To further examine the model's ability to generalize to different tasks on unseen target HGs, we perform node classification and link prediction tasks across four different datasets in zero-shot scenarios. We also fine-tune the model under the \textit{cross-domain and multi-task} setting, followed by evaluation of node classification and link prediction tasks on the unseen target HG.

The results are summarized in Table \ref{tab:main cd-ct zero-shot}. (1) Our MLM4HG outperforms baselines across the majority of datasets and tasks, demonstrating strong generalization capabilities across different graph tasks on unseen target HGs. While it initially exhibits suboptimal and random performance on the node classification task in DBLP, MLM4HG rapidly adapts to the domain of the target HG after \textit{1}-shot fine-tuning in Table \ref{tab:main cd-ct few-shot}, leading to notable improvements in performance. (2) In comparison to the baselines, MLM4HG achieves a significant enhancement in link prediction, with average gains of 27.81\% in Micro-F1 and 13.57\% in Macro-F1. This highlights the effectiveness of metapath-based sequences in enhancing the LM's ability to capture intricate structural information of HGs, leading to significant improvements in link prediction performance.

\begin{table}[t]
    \centering
    % \scriptsize
    \caption{Average Micro-F1 and Macro-F1 scores for node classification in few-shot scenarios. The best results are in bold and the second best are underlined. Our results are significantly better than the best baseline, as confirmed by the Student’s \textit{t}-test at the 0.01 significance level.}
    \resizebox{\textwidth}{!}{
    \setlength{\tabcolsep}{0.9mm}
    % \fontsize{9}{10}
    \begin{tabular}{c|c|c|c|c|c|c|c||c|c|c|c|c|c}
    \toprule
    \multicolumn{2}{c|}{\textbf{Metrix}} & \multicolumn{6}{c||}{\textbf{Micro-F1}} & \multicolumn{6}{c}{\textbf{Macro-F1}} \\
    \midrule
    \textbf{Target} & \textbf{\textit{k}-shot} & \textbf{HAN} & \textbf{HeCo} & \textbf{WalkLM} & \textbf{GraphGPT} & \textbf{HiGPT} & \textbf{MLM4HG} & \textbf{HAN} & \textbf{HeCo} & \textbf{WalkLM} & \textbf{GraphGPT} & \textbf{HiGPT} & \textbf{MLM4HG} \\
    \midrule
    \multirow{3}{*}{\textbf{IMDB}} & \textit{1}-shot &0.3521&0.3401&0.3871&0.3671&\underline{0.4901}&\textbf{0.6529}&0.3126&0.3027&0.3507&0.2782&\underline{0.3621}&\textbf{0.3973}\\
    % \multirow{3}{*}{\textbf{IMDB}} & \textit{1}-shot &0.3521+0.0000&0.3521+0.0000&0.3521+0.0000&0.3671+0.0000&0.4901+0.0000&0.6529+0.0000&0.3126+0.0000&0.3027+0.0000&0.3507+0.0000&0.2782+0.0000&0.3621+0.0000&0.3973+0.0000\\
    & \textit{5}-shot & 0.3622& 0.3423& 0.3701& 0.3996& \underline{0.5849}& \textbf{0.6690}& 0.3117& 0.2977& 0.3448& 0.3305& \underline{0.3817}& \textbf{0.4008}\\
    & \textit{20}-shot & 0.3824& 0.3681& 0.4248& 0.4061& \underline{0.5809}& \textbf{0.7032}& 0.3356& 0.3098& 0.3676& 0.3461& \underline{0.3749}& \textbf{0.4129} \\
    & \textit{40}-shot & 0.4002& 0.3907& 0.4337& 0.4186& \underline{0.6076}& \textbf{0.7201}& 0.3527& 0.3217& 0.4121& 0.3591& \underline{0.3976}& \textbf{0.4186}\\
    \midrule
     
    \multirow{3}{*}{\textbf{DBLP}} & \textit{1}-shot & 0.2538& 0.2401& 0.2542& \underline{0.2831}& 0.2622& \textbf{0.2870}& 0.2435& 0.2398& 0.2452& \underline{0.2497}& 0.2437& \textbf{0.2530}\\
    & \textit{5}-shot & 0.2622& 0.2537& 0.2634& 0.3152& \underline{0.3743}& \textbf{0.3925}& 0.2462& 0.2404& 0.2574& 0.2563& \underline{0.2754}& \textbf{0.2819}\\
    & \textit{20}-shot & 0.2839& 0.2673& 0.3093& 0.3277& \underline{0.3674}& \textbf{0.6350}& 0.2680& 0.2551& 0.2988& 0.3171& \underline{0.3379}& \textbf{0.3884}\\
    & \textit{40}-shot & 0.2903& 0.2894& 0.3176& 0.3301& \underline{0.3824}& \textbf{0.6817}& 0.2742& 0.2635& 0.2799& 0.3205& \underline{0.3577}& \textbf{0.4054}\\
    \midrule
    
    \multirow{3}{*}{\textbf{PubMed}} & \textit{1}-shot & 0.1527& 0.1624& 0.1763& 0.1916& \underline{0.1932}& \textbf{0.5661}& 0.1224& 0.1312& 0.1313& \underline{0.1323}& 0.1268& \textbf{0.3615}\\
    & \textit{5}-shot & 0.2576& 0.2793& \underline{0.3495}& 0.2231& 0.2659& \textbf{0.6203}& 0.1946& 0.2051& \underline{0.2284}& 0.1737& 0.1743& \textbf{0.3828}\\
    & \textit{20}-shot & 0.2603& 0.2832& \underline{0.3593}& 0.2265& 0.2754& \textbf{0.6508}& 0.2275& 0.2307& \underline{0.2382}& 0.1731& 0.1696& \textbf{0.3943}\\
    & \textit{40}-shot & 0.2834& 0.2931& \underline{0.3659}& 0.2438& 0.3037& \textbf{0.6712}& 0.2423& 0.2496& \underline{0.2662}& 0.1933& 0.1807& \textbf{0.4016}\\
    \midrule
    
    \multirow{3}{*}{\textbf{YELP}} & \textit{1}-shot & 0.1427& 0.1135& 0.1613& 0.3164& \underline{0.4133}& \textbf{0.4353}& 0.0981& 0.0879& 0.1242& 0.1856& \underline{0.2833}& \textbf{0.3033}\\
    & \textit{5}-shot & 0.1832& 0.1427& 0.2145& 0.3628& \underline{0.5413}& \textbf{0.6755}& 0.1513& 0.1203& 0.2069& 0.2271& \underline{0.3431}& \textbf{0.4032}\\
    & \textit{20}-shot & 0.2746& 0.2294& 0.2915& 0.3836& \underline{0.5452}& \textbf{0.7524}& 0.2109& 0.1632& 0.2314& 0.2336& \underline{0.3486}& \textbf{0.4294}\\
    & \textit{40}-shot & 0.2831& 0.2475& 0.3023& 0.3906& \underline{0.5713}& \textbf{0.7892}& 0.2379& 0.2070& 0.2401& 0.2765& \underline{0.3624}& \textbf{0.4411}\\
    \bottomrule
    \end{tabular}
    }
    \label{tab:main cd-ct few-shot}
\end{table}

\begin{table}[t]
    \centering
    % \scriptsize
    \caption{Average Micro-F1 and Macro-F1 scores for node classification (NC) and link prediction (LP) in zero-shot scenarios. The best results are in bold and the second best are underlined.}
    \resizebox{\textwidth}{!}{
    \setlength{\tabcolsep}{0.9mm}
    \begin{tabular}{c|c|c|c|c|c|c|c||c|c|c|c|c|c}
    \toprule
    
    \multicolumn{2}{c|}{\textbf{Metric}} & \multicolumn{6}{c||}{\textbf{Micro-F1}} & \multicolumn{6}{c}{\textbf{Macro-F1}} \\
    \midrule
    
    \textbf{Task} & \textbf{Target} & \textbf{HAN} & \textbf{HeCo} & \textbf{WalkLM} & \textbf{GraphGPT} & \textbf{HiGPT} & \textbf{MLM4HG} & \textbf{HAN} & \textbf{HeCo} & \textbf{WalkLM} & \textbf{GraphGPT} & \textbf{HiGPT} & \textbf{MLM4HG} \\
    \midrule
    
    \multirow{4}{*}{\textbf{NC}} & \textbf{IMDB} & 0.3233& 0.3203& 0.3456& 0.3525& \underline{0.3895}& \textbf{0.4622}& 0.2201& 0.2275& 0.2352& 0.2411& \underline{0.2838}& \textbf{0.3161}\\ 
    & \textbf{DBLP} & 0.2405& 0.2437& \textbf{0.2483}& 0.2465& 0.2371& \underline{0.2477}& 0.1594& 0.1600& \underline{0.1638}& \textbf{0.1659}& 0.1532& 0.1601\\
    & \textbf{PubMed} & 0.1435& 0.1324& 0.1536& 0.1601& \underline{0.1811}& \textbf{0.1898}& 0.0827& 0.0601& 0.0514& 0.1032& \underline{0.1232}& \textbf{0.1595}\\
    & \textbf{YELP} & 0.0635& 0.0721& 0.0798& 0.1138& \underline{0.1532}& \textbf{0.1772}& 0.0273& 0.0298& 0.0399& 0.0772& \underline{0.1134}& \textbf{0.1505}\\
    \midrule
    
    \multirow{4}{*}{\textbf{LP}} & \textbf{IMDB} & 0.1625& 0.1542& 0.1930& 0.2066& \underline{0.3662}& \textbf{0.9054}& 0.0814& 0.0793& 0.0816& 0.1176& \underline{0.1868}& \textbf{0.4752}\\ 
    & \textbf{DBLP} & 0.1236& 0.1149& 0.1397& 0.2923& \underline{0.3246}& \textbf{0.5356}& 0.0779& 0.0683& 0.0843& \underline{0.2403}& 0.2355& \textbf{0.3488}\\
    & \textbf{PubMed} & 0.1764& 0.1527& 0.1830& 0.1866& \underline{0.1912}& \textbf{0.1986}& 0.0825& 0.0673& 0.0948& 0.1125& \underline{0.1352}& \textbf{0.1657}\\
    & \textbf{YELP} & 0.1301& 0.1232& 0.1456& 0.2491& \underline{0.4829}& \textbf{0.8378}& 0.0649& 0.0587& 0.0766& 0.0881& \underline{0.3404}& \textbf{0.4559}\\
    \bottomrule
    
    \end{tabular}
    }
    \label{tab:main cd-ct zero-shot}
\end{table}

\subsection{Ablation Study (RQ3)}
In this subsection, we introduce four variants to examine the influence of different learning scenarios and model components on the experimental results. To capture more comprehensive and general graph knowledge from source HGs, MLM4HG adopts \textit{cross-domain and multi-task} fine-tuning by default. To evaluate whether the model benefits from this learning scenarios, we define two variants: \textbf{w/o Fine-tuning} and \textbf{w/o Multi-task}. \textbf{w/o Fine-tuning} indicates that the model is directly applied to the target HG without \textit{cross-domain and multi-task} fine-tuning on source HGs. \textbf{w/o Multi-task} refers to fine-tuning the pretrained LM on source HGs using the single same task as the target HG. That is, \textbf{w/o Multi-task} adds multi-domain pretraining to \textbf{w/o Fine-tuning}, whereas the complete MLM4HG includes both multi-domain and multi-task pretraining. Additionally, to assess the impact of different components on the experimental outcomes, we design two variants: \textbf{w/o Metapath} and \textbf{w/o Constraint}. In \textbf{w/o Metapath}, the metapath-based sequences is replaced with node attributes. In \textbf{w/o Constraint}, the predicted \texttt{<mask>} token is selected from the full vocabulary rather than the constrained target vocabulary.

We evaluated the performance of these variants in node classification and link prediction tasks in both few-shot and zero-shot scenarios. The experimental results are shown in Table \ref{tab:different setting evaluations}. (1) It is evident that these variants perform suboptimally compared to the original MLM4HG across different graph tasks and experimental settings. (2) \textbf{w/o Multi-task} outperforms \textbf{w/o Fine-tuning} and achieves competitive results overall, demonstrating that multi-domain pretraining enables the LM to extract transferable knowledge and significantly boosts performance, even when fine-tuned on a single task. Further performance gains achieved by MLM4HG highlight the benefit of multi-task fine-tuning, which facilitates the extraction of more comprehensive information from multiple perspectives. (3) When the vocabulary size is unconstrained in \textbf{w/o Constraint}, the results experience a significant decline compared to MLM4HG. This clearly indicates that the model’s focus should be directed toward the target categories, especially for our simpler graph tasks. (4) When node attributes replace metapath-based sequences in \textbf{w/o Metapath}, the performance of both node classification and link prediction tasks also decreases to varying degrees. This confirms that metapath-based sequences contains rich structural and semantic information, which helps the LM extract general heterogeneous graph knowledge and generalizing across different HGs. (5) We further observe that in the few-shot scenarios, all variants—except for w/o Constraint—perform comparably to MLM4HG. This is because the pretrained LM can quickly adapt to a small number of labeled examples, as shown in Table \ref{tab:main cd-ct few-shot}. For instance, our methods already outperform the strongest baselines under the 5-shot setting on the IMDB and YELP datasets. However, in the more challenging zero-shot scenario, each component becomes essential.

\begin{table}[tbp]
    \centering
    % \scriptsize
    \caption{Average Micro-F1 and Macro-F1 scores for MLM4HG variants on node classification and link prediction tasks over IMDB and YELP datasets.}
    \resizebox{\textwidth}{!}{
    \begin{tabular}{c|c|c|c|c|c|c|c|c|c|c|c|c}
    \toprule
    
    \textbf{Task} & \multicolumn{8}{c|}{\textbf{Node Classification}} & \multicolumn{4}{c}{\textbf{Link Prediction}} \\
    \midrule
    
    \textbf{\textit{k}-shot} & \multicolumn{4}{c|}{\textbf{\textit{5}-shot}} & \multicolumn{4}{c|}{\textbf{\textit{zero}-shot}} & \multicolumn{4}{c}{\textbf{\textit{zero}-shot}} \\
    \midrule

    \textbf{Metric} & \multicolumn{2}{c|}{\textbf{Micro-F1}} & \multicolumn{2}{c|}{\textbf{Macro-F1}} & \multicolumn{2}{c|}{\textbf{Micro-F1}} & \multicolumn{2}{c|}{\textbf{Macro-F1}} & \multicolumn{2}{c|}{\textbf{Micro-F1}} & \multicolumn{2}{c}{\textbf{Macro-F1}} \\
    \midrule

    \textbf{Target} & \textbf{IMDB} & \textbf{YELP} & \textbf{IMDB} & \textbf{YELP}& \textbf{IMDB} & \textbf{YELP}& \textbf{IMDB} & \textbf{YELP}& \textbf{IMDB} & \textbf{YELP}& \textbf{IMDB} & \textbf{YELP} \\
    \midrule

    \textbf{w/o Fine-tuning} & 0.6497& 0.6604& 0.3875& 0.3977& 0.3666&0.0746 &0.2682 &0.0695 &0.5254 &0.1781 &0.3444 &0.1511 \\
    \textbf{w/o Multi-task} & \underline{0.6570}& \underline{0.6701}& \underline{0.3965}& \underline{0.4012}&\underline{0.4598} &0.1476 &\underline{0.3150} &0.1286 &\underline{0.8853} &\underline{0.8162} &\underline{0.4696} & \underline{0.4494}\\
    \textbf{w/o Metapath} & 0.6380& 0.4917& 0.3795& 0.3296& 0.4304& \underline{0.1519}& 0.2867& \underline{0.1389}& 0.8705& 0.3173& 0.4654& 0.2409\\
    \textbf{w/o Constraint} & 0.3643& 0.0547& 0.2670& 0.0519& 0.0491& 0.0000& 0.0468& 0.0000& 0.4302& 0.1772& 0.3008& 0.1505\\
    \textbf{MLM4HG} & \textbf{0.6690}& \textbf{0.6755}& \textbf{0.4008}& \textbf{0.4032}& \textbf{0.4622}& \textbf{0.1772}& \textbf{0.3161}& \textbf{0.1505}& \textbf{0.9054}& \textbf{0.8378}& \textbf{0.4752}& \textbf{0.4559}\\
    \bottomrule
    \end{tabular}
    }
    \label{tab:different setting evaluations}
\end{table}

\subsection{Graph-to-Text Method Comparison (RQ4)}
We compare our metapath-based sequences with commonly used graph-to-text methods, specifically GML \cite{jinyu:14} and GraphML \cite{jinyu:15}. Networkx provides a unified parser for these two formats, thus we use the parsed node and edge information as a replacement for the metapath-based sequences. Examples and parsing results of these two formats are provided in Appendix \ref{app:examples and parsing results}.

% As shown in Table 4, the metapath-based sequences demonstrate superior performance in both 326
% node classification and link prediction tasks on the IMDB and YELP datasets. Compared to our 327
% metapath-based sequences, GML and GraphML merely offer a straightforward description of nodes 328
% and edges, requiring the LM to further extract the rich and implicit semantic information, which 329
% increases the task complexity. This confirms that our metapath-based sequences effectively extracts 330
% both structural and semantic information from HGs, benefiting downstream tasks.

As shown in Table \ref{tab:different graph-to-text methods}, metapath-based sequences demonstrate superior performance in both node classification and link prediction tasks on the IMDB and YELP datasets. Compared to metapath-based sequences, GML and GraphML merely offer a straightforward description of nodes and edges, requiring the LM to further extract the rich and implicit semantic information, which increases the task complexity. This confirms that our metapath-based sequences effectively extracts both structural and semantic information from HGs, benefiting downstream tasks.

\begin{table}[htbp]
    \centering
    % \scriptsize
    \caption{Average Micro-F1 and Macro-F1 scores for node classification and link prediction using different graph-to-text methods on IMDB and YELP under both few-shot and zero-shot scenarios.}
    \resizebox{\textwidth}{!}{
    \begin{tabular}{c|c|c|c|c|c|c|c|c|c|c|c|c}
    \toprule
    
    \textbf{Task} & \multicolumn{8}{c|}{\textbf{Node Classification}} & \multicolumn{4}{c}{\textbf{Link Prediction}} \\
    \midrule

    \textbf{\textit{k}-shot} & \multicolumn{4}{c|}{\textbf{\textit{5}-shot}} & \multicolumn{4}{c|}{\textbf{\textit{zero}-shot}} & \multicolumn{4}{c}{\textbf{\textit{zero}-shot}} \\
    \midrule

    \textbf{Metric} & \multicolumn{2}{c|}{\textbf{Micro-F1}} & 
    \multicolumn{2}{c|}{\textbf{Macro-F1}} & \multicolumn{2}{c|}{\textbf{Micro-F1}} & \multicolumn{2}{c|}{\textbf{Macro-F1}} & \multicolumn{2}{c|}{\textbf{Micro-F1}} & \multicolumn{2}{c}{\textbf{Macro-F1}} \\
    \midrule

    \textbf{Target} & \textbf{IMDB} & \textbf{YELP} & \textbf{IMDB} & \textbf{YELP} & \textbf{IMDB} & \textbf{YELP} & \textbf{IMDB} & \textbf{YELP} & \textbf{IMDB} & \textbf{YELP} & \textbf{IMDB} & \textbf{YELP} \\
    \midrule
    
    \textbf{GML \& GraphML} & 0.6337& 0.5764& 0.3879& 0.3656& 0.4043& 0.0701& 0.2879& 0.0665& 0.8984& 0.6406&0.4587 &0.3059 \\
    \textbf{Metapath-based sequences} & \textbf{0.6690}& \textbf{0.6755}& \textbf{0.4008}& \textbf{0.4032}& \textbf{0.4622}& \textbf{0.1772}& \textbf{0.3161}& \textbf{0.1505}& \textbf{0.9054}& \textbf{0.8378}& \textbf{0.4762}& \textbf{0.4559}\\
    \bottomrule
    
    \end{tabular}
    }
    \label{tab:different graph-to-text methods}
\end{table}

\section{Conclusion}
\label{main:conclusion}
In this paper, we propose a simple yet effective Masked Language Modeling-based approach for generalizable Heterogeneous Graph learning, called MLM4HG. MLM4HG introduces metapath-based sequences to unify the representation of HGs, enabling cross-domain generalization. Then it reformulates different tasks into a cloze-style \texttt{<mask>} token prediction paradigm using customized templates, thereby supporting multi-task generalization. By constructing an HG-based corpus and using it to fine-tune a pretrained LM, we enable the LM to generalize across unseen target graphs and tasks. To the best of our knowledge, we are the first to apply masked language models for generalizable heterogeneous graph learning, establishing a new paradigm and paving a new path for future research. Extensive experiments in four real-world datasets from distinct domains demonstrate MLM4HG's superior performance across different graph tasks in both few-shot and zero-shot scenarios. Limitations and broader impacts are discussed in Appendix~\ref{app:limitation}.

% \noindent\textbf{Limitations.} While MLM4HG effectively captures structural signals from metapath-based sequences (Section~\ref{sec:explore}), the semantic information from textual content remains crucial for LMs. Thus, applying LMs to non-textual graph data still requires further exploration. Moreover, additional investigation is needed to assess the model’s generalizability across a wider range of tasks, such as regression tasks. 

% \noindent\textbf{Broader Impacts.} MLM4HG holds promise for robust AI development in real-world applications involving HGs, including social networks, knowledge graphs, and recommender systems.% However, its generalization ability may introduce risks of misuse and unintended bias. Future work should emphasize transparency and fairness, particularly in high-stakes scenarios.

{
\small

\bibliographystyle{plain}
\bibliography{neurips_2025}

\begin{thebibliography}{48}
\providecommand{\natexlab}[1]{#1}
\providecommand{\url}[1]{\texttt{#1}}
\expandafter\ifx\csname urlstyle\endcsname\relax
  \providecommand{\doi}[1]{doi: #1}\else
  \providecommand{\doi}{doi: \begingroup \urlstyle{rm}\Url}\fi

\bibitem[Achiam et~al.(2023)Achiam, Adler, Agarwal, Ahmad, Akkaya, Aleman, Almeida, Altenschmidt, Altman, Anadkat, et~al.]{jinyu:28}
Achiam, J., Adler, S., Agarwal, S., Ahmad, L., Akkaya, I., Aleman, F.~L., Almeida, D., Altenschmidt, J., Altman, S., Anadkat, S., et~al.
\newblock Gpt-4 technical report.
\newblock \emph{arXiv preprint arXiv:2303.08774}, 2023.

\bibitem[Brandes et~al.(2013)Brandes, Eiglsperger, Lerner, and Pich]{jinyu:15}
Brandes, U., Eiglsperger, M., Lerner, J., and Pich, C.
\newblock Graph markup language (graphml).
\newblock 2013.

\bibitem[Brown et~al.(2020)Brown, Mann, Ryder, Subbiah, Kaplan, Dhariwal, Neelakantan, Shyam, Sastry, Askell, et~al.]{jinyu:36}
Brown, T., Mann, B., Ryder, N., Subbiah, M., Kaplan, J.~D., Dhariwal, P., Neelakantan, A., Shyam, P., Sastry, G., Askell, A., et~al.
\newblock Language models are few-shot learners.
\newblock \emph{Advances in neural information processing systems}, 33:\penalty0 1877--1901, 2020.

\bibitem[Cai et~al.(2024)Cai, Tan, Lei, Zhu, Wang, Zheng, and Luo]{jinyu:1}
Cai, Z., Tan, Z., Lei, Z., Zhu, Z., Wang, H., Zheng, Q., and Luo, M.
\newblock Lmbot: distilling graph knowledge into language model for graph-less deployment in twitter bot detection.
\newblock In \emph{Proceedings of the 17th ACM International Conference on Web Search and Data Mining}, pp.\  57--66, 2024.

\bibitem[Chai et~al.(2023)Chai, Zhang, Wu, Han, Hu, Huang, and Yang]{jinyu:38}
Chai, Z., Zhang, T., Wu, L., Han, K., Hu, X., Huang, X., and Yang, Y.
\newblock Graphllm: Boosting graph reasoning ability of large language model.
\newblock \emph{arXiv preprint arXiv:2310.05845}, 2023.

\bibitem[Chen et~al.(2023)Chen, Huang, Xia, Wei, Xu, and Luo]{jinyu:2}
Chen, M., Huang, C., Xia, L., Wei, W., Xu, Y., and Luo, R.
\newblock Heterogeneous graph contrastive learning for recommendation.
\newblock In \emph{Proceedings of the sixteenth ACM international conference on web search and data mining}, pp.\  544--552, 2023.

\bibitem[Chen et~al.(2024)Chen, Zhao, Jaiswal, Shah, and Wang]{jinyu:12}
Chen, R., Zhao, T., Jaiswal, A., Shah, N., and Wang, Z.
\newblock Llaga: Large language and graph assistant.
\newblock \emph{arXiv preprint arXiv:2402.08170}, 2024.

\bibitem[Devlin(2018)]{jinyu:16}
Devlin, J.
\newblock Bert: Pre-training of deep bidirectional transformers for language understanding.
\newblock \emph{arXiv preprint arXiv:1810.04805}, 2018.

\bibitem[Dong et~al.(2017)Dong, Chawla, and Swami]{jinyu:34}
Dong, Y., Chawla, N.~V., and Swami, A.
\newblock metapath2vec: Scalable representation learning for heterogeneous networks.
\newblock In \emph{Proceedings of the 23rd ACM SIGKDD international conference on knowledge discovery and data mining}, pp.\  135--144, 2017.

\bibitem[Fey \& Lenssen(2019)Fey and Lenssen]{jinyu:22}
Fey, M. and Lenssen, J.~E.
\newblock Fast graph representation learning with pytorch geometric.
\newblock \emph{arXiv preprint arXiv:1903.02428}, 2019.

\bibitem[Fu et~al.(2020)Fu, Zhang, Meng, and King]{jinyu:6}
Fu, X., Zhang, J., Meng, Z., and King, I.
\newblock Magnn: Metapath aggregated graph neural network for heterogeneous graph embedding.
\newblock In \emph{Proceedings of the web conference 2020}, pp.\  2331--2341, 2020.

\bibitem[Himsolt(1997)]{jinyu:14}
Himsolt, M.
\newblock Gml: A portable graph file format.
\newblock Technical report, Technical report, Universitat Passau, 1997.

\bibitem[Hu et~al.(2021)Hu, Shen, Wallis, Allen-Zhu, Li, Wang, Wang, and Chen]{jinyu:49}
Hu, E.~J., Shen, Y., Wallis, P., Allen-Zhu, Z., Li, Y., Wang, S., Wang, L., and Chen, W.
\newblock Lora: Low-rank adaptation of large language models.
\newblock \emph{arXiv preprint arXiv:2106.09685}, 2021.

\bibitem[Hu et~al.(2020{\natexlab{a}})Hu, Dong, Wang, Chang, and Sun]{jinyu:46}
Hu, Z., Dong, Y., Wang, K., Chang, K.-W., and Sun, Y.
\newblock Gpt-gnn: Generative pre-training of graph neural networks.
\newblock In \emph{Proceedings of the 26th ACM SIGKDD international conference on knowledge discovery \& data mining}, pp.\  1857--1867, 2020{\natexlab{a}}.

\bibitem[Hu et~al.(2020{\natexlab{b}})Hu, Dong, Wang, and Sun]{jinyu:33}
Hu, Z., Dong, Y., Wang, K., and Sun, Y.
\newblock Heterogeneous graph transformer.
\newblock In \emph{Proceedings of the web conference 2020}, pp.\  2704--2710, 2020{\natexlab{b}}.

\bibitem[Jiang et~al.(2021)Jiang, Jia, Fang, Shi, Lin, and Wang]{jinyu:47}
Jiang, X., Jia, T., Fang, Y., Shi, C., Lin, Z., and Wang, H.
\newblock Pre-training on large-scale heterogeneous graph.
\newblock In \emph{Proceedings of the 27th ACM SIGKDD conference on knowledge discovery \& data mining}, pp.\  756--766, 2021.

\bibitem[Li et~al.(2024)Li, Wang, Li, Yu, and Li]{jinyu:43}
Li, Y., Wang, P., Li, Z., Yu, J.~X., and Li, J.
\newblock Zerog: Investigating cross-dataset zero-shot transferability in graphs.
\newblock In \emph{Proceedings of the 30th ACM SIGKDD Conference on Knowledge Discovery and Data Mining}, pp.\  1725--1735, 2024.

\bibitem[Lin et~al.(2024)Lin, Yan, Song, Jiang, Kang, Lin, Yuan, Cao, Sun, and Liu]{jinyu:13}
Lin, T., Yan, P., Song, K., Jiang, Z., Kang, Y., Lin, J., Yuan, W., Cao, J., Sun, C., and Liu, X.
\newblock Langgfm: A large language model alone can be a powerful graph foundation model.
\newblock \emph{arXiv preprint arXiv:2410.14961}, 2024.

\bibitem[Liu et~al.(2023{\natexlab{a}})Liu, Feng, Kong, Liang, Tao, Chen, and Zhang]{jinyu:41}
Liu, H., Feng, J., Kong, L., Liang, N., Tao, D., Chen, Y., and Zhang, M.
\newblock One for all: Towards training one graph model for all classification tasks.
\newblock \emph{arXiv preprint arXiv:2310.00149}, 2023{\natexlab{a}}.

\bibitem[Liu et~al.(2023{\natexlab{b}})Liu, Yang, Lu, Chen, Li, Zhang, Bai, Fang, Sun, Yu, et~al.]{jinyu:37}
Liu, J., Yang, C., Lu, Z., Chen, J., Li, Y., Zhang, M., Bai, T., Fang, Y., Sun, L., Yu, P.~S., et~al.
\newblock Towards graph foundation models: A survey and beyond.
\newblock \emph{arXiv preprint arXiv:2310.11829}, 2023{\natexlab{b}}.

\bibitem[Liu(2019)]{jinyu:18}
Liu, Y.
\newblock Roberta: A robustly optimized bert pretraining approach.
\newblock \emph{arXiv preprint arXiv:1907.11692}, 364, 2019.

\bibitem[Liu et~al.(2024)Liu, He, Tian, and Chawla]{jinyu:39}
Liu, Z., He, X., Tian, Y., and Chawla, N.~V.
\newblock Can we soft prompt llms for graph learning tasks?
\newblock In \emph{Companion Proceedings of the ACM on Web Conference 2024}, pp.\  481--484, 2024.

\bibitem[Loshchilov(2017)]{jinyu:26}
Loshchilov, I.
\newblock Decoupled weight decay regularization.
\newblock \emph{arXiv preprint arXiv:1711.05101}, 2017.

\bibitem[Lu et~al.(2019)Lu, Shi, Hu, and Liu]{jinyu:31}
Lu, Y., Shi, C., Hu, L., and Liu, Z.
\newblock Relation structure-aware heterogeneous information network embedding.
\newblock In \emph{Proceedings of the AAAI conference on artificial intelligence}, volume~33, pp.\  4456--4463, 2019.

\bibitem[Ma et~al.(2023)Ma, Wang, Li, Wang, Xiao, Liu, Cheng, Wang, Li, Chang, et~al.]{jinyu:4}
Ma, A., Wang, X., Li, J., Wang, C., Xiao, T., Liu, Y., Cheng, H., Wang, J., Li, Y., Chang, Y., et~al.
\newblock Single-cell biological network inference using a heterogeneous graph transformer.
\newblock \emph{Nature Communications}, 14\penalty0 (1):\penalty0 964, 2023.

\bibitem[Paszke et~al.(2019)Paszke, Gross, Massa, Lerer, Bradbury, Chanan, Killeen, Lin, Gimelshein, Antiga, et~al.]{jinyu:21}
Paszke, A., Gross, S., Massa, F., Lerer, A., Bradbury, J., Chanan, G., Killeen, T., Lin, Z., Gimelshein, N., Antiga, L., et~al.
\newblock Pytorch: An imperative style, high-performance deep learning library.
\newblock \emph{Advances in neural information processing systems}, 32, 2019.

\bibitem[Pedregosa et~al.(2011)Pedregosa, Varoquaux, Gramfort, Michel, Thirion, Grisel, Blondel, Prettenhofer, Weiss, Dubourg, et~al.]{jinyu:24}
Pedregosa, F., Varoquaux, G., Gramfort, A., Michel, V., Thirion, B., Grisel, O., Blondel, M., Prettenhofer, P., Weiss, R., Dubourg, V., et~al.
\newblock Scikit-learn: Machine learning in python.
\newblock \emph{the Journal of machine Learning research}, 12:\penalty0 2825--2830, 2011.

\bibitem[Sanh(2019)]{jinyu:17}
Sanh, V.
\newblock Distilbert, a distilled version of bert: smaller, faster, cheaper and lighter.
\newblock \emph{arXiv preprint arXiv:1910.01108}, 2019.

\bibitem[Schlichtkrull et~al.(2018)Schlichtkrull, Kipf, Bloem, Van Den~Berg, Titov, and Welling]{jinyu:32}
Schlichtkrull, M., Kipf, T.~N., Bloem, P., Van Den~Berg, R., Titov, I., and Welling, M.
\newblock Modeling relational data with graph convolutional networks.
\newblock In \emph{The semantic web: 15th international conference, ESWC 2018, Heraklion, Crete, Greece, June 3--7, 2018, proceedings 15}, pp.\  593--607. Springer, 2018.

\bibitem[Sun et~al.(2019)Sun, Deng, Nie, and Tang]{jinyu:3}
Sun, Z., Deng, Z.-H., Nie, J.-Y., and Tang, J.
\newblock Rotate: Knowledge graph embedding by relational rotation in complex space.
\newblock \emph{arXiv preprint arXiv:1902.10197}, 2019.

\bibitem[Tan et~al.(2024)Tan, Zhou, Lv, Liu, and Yang]{jinyu:19}
Tan, Y., Zhou, Z., Lv, H., Liu, W., and Yang, C.
\newblock Walklm: A uniform language model fine-tuning framework for attributed graph embedding.
\newblock \emph{Advances in Neural Information Processing Systems}, 36, 2024.

\bibitem[Tang et~al.(2024{\natexlab{a}})Tang, Yang, Wei, Shi, Su, Cheng, Yin, and Huang]{jinyu:11}
Tang, J., Yang, Y., Wei, W., Shi, L., Su, L., Cheng, S., Yin, D., and Huang, C.
\newblock Graphgpt: Graph instruction tuning for large language models.
\newblock In \emph{Proceedings of the 47th International ACM SIGIR Conference on Research and Development in Information Retrieval}, pp.\  491--500, 2024{\natexlab{a}}.

\bibitem[Tang et~al.(2024{\natexlab{b}})Tang, Yang, Wei, Shi, Xia, Yin, and Huang]{jinyu:10}
Tang, J., Yang, Y., Wei, W., Shi, L., Xia, L., Yin, D., and Huang, C.
\newblock Higpt: Heterogeneous graph language model.
\newblock In \emph{Proceedings of the 30th ACM SIGKDD Conference on Knowledge Discovery and Data Mining}, pp.\  2842--2853, 2024{\natexlab{b}}.

\bibitem[Tian et~al.(2023)Tian, Dong, Zhang, Zhang, and Chawla]{jinyu:48}
Tian, Y., Dong, K., Zhang, C., Zhang, C., and Chawla, N.~V.
\newblock Heterogeneous graph masked autoencoders.
\newblock In \emph{Proceedings of the AAAI Conference on Artificial Intelligence}, volume~37, pp.\  9997--10005, 2023.

\bibitem[Touvron et~al.(2023)Touvron, Lavril, Izacard, Martinet, Lachaux, Lacroix, Rozi{\`e}re, Goyal, Hambro, Azhar, et~al.]{jinyu:27}
Touvron, H., Lavril, T., Izacard, G., Martinet, X., Lachaux, M.-A., Lacroix, T., Rozi{\`e}re, B., Goyal, N., Hambro, E., Azhar, F., et~al.
\newblock Llama: open and efficient foundation language models. arxiv.
\newblock \emph{arXiv preprint arXiv:2302.13971}, 2023.

\bibitem[Wang et~al.(2024)Wang, Feng, He, Tan, Han, and Tsvetkov]{jinyu:29}
Wang, H., Feng, S., He, T., Tan, Z., Han, X., and Tsvetkov, Y.
\newblock Can language models solve graph problems in natural language?
\newblock \emph{Advances in Neural Information Processing Systems}, 36, 2024.

\bibitem[Wang et~al.(2019{\natexlab{a}})Wang, Zheng, Ye, Gan, Li, Song, Zhou, Ma, Yu, Gai, et~al.]{jinyu:23}
Wang, M., Zheng, D., Ye, Z., Gan, Q., Li, M., Song, X., Zhou, J., Ma, C., Yu, L., Gai, Y., et~al.
\newblock Deep graph library: A graph-centric, highly-performant package for graph neural networks.
\newblock \emph{arXiv preprint arXiv:1909.01315}, 2019{\natexlab{a}}.

\bibitem[Wang et~al.(2019{\natexlab{b}})Wang, Ji, Shi, Wang, Ye, Cui, and Yu]{jinyu:5}
Wang, X., Ji, H., Shi, C., Wang, B., Ye, Y., Cui, P., and Yu, P.~S.
\newblock Heterogeneous graph attention network.
\newblock In \emph{The world wide web conference}, pp.\  2022--2032, 2019{\natexlab{b}}.

\bibitem[Wang et~al.(2021)Wang, Liu, Han, and Shi]{jinyu:45}
Wang, X., Liu, N., Han, H., and Shi, C.
\newblock Self-supervised heterogeneous graph neural network with co-contrastive learning.
\newblock In \emph{Proceedings of the 27th ACM SIGKDD conference on knowledge discovery \& data mining}, pp.\  1726--1736, 2021.

\bibitem[Wang et~al.(2022)Wang, Bo, Shi, Fan, Ye, and Philip]{jinyu:30}
Wang, X., Bo, D., Shi, C., Fan, S., Ye, Y., and Philip, S.~Y.
\newblock A survey on heterogeneous graph embedding: methods, techniques, applications and sources.
\newblock \emph{IEEE Transactions on Big Data}, 9\penalty0 (2):\penalty0 415--436, 2022.

\bibitem[Wolf et~al.(2020)Wolf, Debut, Sanh, Chaumond, Delangue, Moi, Cistac, Rault, Louf, Funtowicz, et~al.]{jinyu:25}
Wolf, T., Debut, L., Sanh, V., Chaumond, J., Delangue, C., Moi, A., Cistac, P., Rault, T., Louf, R., Funtowicz, M., et~al.
\newblock Transformers: State-of-the-art natural language processing.
\newblock In \emph{Proceedings of the 2020 conference on empirical methods in natural language processing: system demonstrations}, pp.\  38--45, 2020.

\bibitem[Yang et~al.(2025)Yang, Li, Yang, Zhang, Hui, Zheng, Yu, Gao, Huang, Lv, Zheng, Liu, Zhou, Huang, Hu, Ge, Wei, Lin, Tang, Yang, Tu, Zhang, Yang, Yang, Zhou, Zhou, Lin, Dang, Bao, Yang, Yu, Deng, Li, Xue, Li, Zhang, Wang, Zhu, Men, Gao, Liu, Luo, Li, Tang, Yin, Ren, Wang, Zhang, Ren, Fan, Su, Zhang, Zhang, Wan, Liu, Wang, Cui, Zhang, Zhou, and Qiu]{jinyu:56}
Yang, A., Li, A., Yang, B., Zhang, B., Hui, B., Zheng, B., Yu, B., Gao, C., Huang, C., Lv, C., Zheng, C., Liu, D., Zhou, F., Huang, F., Hu, F., Ge, H., Wei, H., Lin, H., Tang, J., Yang, J., Tu, J., Zhang, J., Yang, J., Yang, J., Zhou, J., Zhou, J., Lin, J., Dang, K., Bao, K., Yang, K., Yu, L., Deng, L., Li, M., Xue, M., Li, M., Zhang, P., Wang, P., Zhu, Q., Men, R., Gao, R., Liu, S., Luo, S., Li, T., Tang, T., Yin, W., Ren, X., Wang, X., Zhang, X., Ren, X., Fan, Y., Su, Y., Zhang, Y., Zhang, Y., Wan, Y., Liu, Y., Wang, Z., Cui, Z., Zhang, Z., Zhou, Z., and Qiu, Z.
\newblock Qwen3 technical report.
\newblock \emph{arXiv preprint arXiv:2505.09388}, 2025.

\bibitem[Yang et~al.(2020)Yang, Xiao, Zhang, Sun, and Han]{jinyu:9}
Yang, C., Xiao, Y., Zhang, Y., Sun, Y., and Han, J.
\newblock Heterogeneous network representation learning: A unified framework with survey and benchmark.
\newblock \emph{IEEE Transactions on Knowledge and Data Engineering}, 34\penalty0 (10):\penalty0 4854--4873, 2020.

\bibitem[Yang et~al.(2021)Yang, Guan, Li, Zhao, Cui, and Wang]{jinyu:51}
Yang, Y., Guan, Z., Li, J., Zhao, W., Cui, J., and Wang, Q.
\newblock Interpretable and efficient heterogeneous graph convolutional network.
\newblock \emph{IEEE Transactions on Knowledge and Data Engineering}, 35\penalty0 (2):\penalty0 1637--1650, 2021.

\bibitem[Yang et~al.(2022)Yang, Guan, Wang, Zhao, Xu, Lu, and Huang]{jinyu:52}
Yang, Y., Guan, Z., Wang, Z., Zhao, W., Xu, C., Lu, W., and Huang, J.
\newblock Self-supervised heterogeneous graph pre-training based on structural clustering.
\newblock \emph{Advances in Neural Information Processing Systems}, 35:\penalty0 16962--16974, 2022.

\bibitem[Ye et~al.(2023)Ye, Zhang, Wang, Xu, Zhang, et~al.]{jinyu:42}
Ye, R., Zhang, C., Wang, R., Xu, S., Zhang, Y., et~al.
\newblock Natural language is all a graph needs.
\newblock \emph{arXiv preprint arXiv:2308.07134}, 4\penalty0 (5):\penalty0 7, 2023.

\bibitem[Zhang et~al.(2024)Zhang, Du, Huang, Wang, and Wang]{jinyu:54}
Zhang, Q., Du, T., Huang, H., Wang, Y., and Wang, Y.
\newblock Look ahead or look around? a theoretical comparison between autoregressive and masked pretraining.
\newblock \emph{arXiv preprint arXiv:2407.00935}, 2024.

\bibitem[Zhu et~al.()Zhu, Xue, Zhao, Jin, Xu, Huang, Wang, Zhou, and Zhang]{jinyu:40}
Zhu, X., Xue, H., Zhao, Z., Jin, M., Xu, W., Huang, J., Wang, Q., Zhou, K., and Zhang, Y.
\newblock Llm as gnn: Graph vocabulary learning for graph foundation model.

\end{thebibliography}
}

%%%%%%%%%%%%%%%%%%%%%%%%%%%%%%%%%%%%%%%%%%%%%%%%%%%%%%%%%%%%
\newpage
\appendix

\section{Dataset Statistics}
\label{app:dataset}
Table \ref{tab:dataset_statistics} summarizes dataset statistics, including node/edge types (with the target node type in bold), target categories, and metapaths used in the metapath-based sequence construction.

\begin{table*}[htbp]
	\centering 
    % \scriptsize
    \caption{Statistics of datasets.}
    \resizebox{\textwidth}{!}{
	\begin{tabular}{ccccc}
		\toprule
		\textbf{Dataset} & \textbf{Node Type} & \textbf{Edge Type} & \textbf{Target Node Category} & \textbf{Metapath} \\
		\midrule
		IMDB 
        & \makecell[c]{\textbf{Movie (M)}, 4,278\\Actor (A), 5,257\\Director (D), 2,081} 
        & \makecell[c]{acted in, 12,828\\was acted by, 12,828\\directed, 4,278\\was directed by, 4,278} 
        & \makecell[c]{Action\\Comedy\\Drama} 
        & \makecell[c]{MAM, MDM\\AMA, AMDMA\\DMD, DMAMD}  \\
		\midrule
		DBLP 
        & \makecell[c]{\textbf{Author (A)}, 4,057\\Paper (P), 14,328\\Term (T), 7,723\\Conference (C), 20} 
        & \makecell[c]{write, 19,645\\was written by, 19,645\\publish, 85,810\\was published in, 85,810\\receive, 14,328\\was received by, 14,328}
        & \makecell[c]{1, 2, 3, 4}
        & \makecell[c]{APA, APCPA, APTPA\\PAP, PTP, PCP\\TPT, TPCPT, TPAPT\\CPC, CPAPC, CPTPC} \\
        \midrule
		YELP 
        & \makecell[c]{\textbf{Business (B)}, 7,474\\Location (L), 39\\Phrase (P), 74,943\\Stars (S), 9}
        & \makecell[c]{rate, 7,474\\score, 7,474\\indicate, 2,654,313\\located in, 7,474\\described with, 2,654,313\\is associated with, 7,474}
        & \makecell[c]{beauty \& spas\\pizza, food\\sandwiches\\nightlife\\burgers\\Mexican\\American (new)\\shopping\\automotive\\Italian, bars\\restaurants\\breakfast \& brunch\\American (traditional)\\event planning \& services}
        & \makecell[c]{BLB, BSB, BLB\\LBL, LBPBL, LBSBL\\SBS, SBLBS, SBPBS\\PBP, PBLBP, PBSBP} \\
        \midrule
		PubMed 
        & \makecell[c]{\textbf{Disease (D)}, 20,163\\Gene (G), 13,562\\Chemical (C), 26,522\\Species (S), 2,863} 
        & \makecell[c]{in, 97,297\\with, 8,400\\contain, 88,897\\causing, 25,962\\is caused by, 25,962}
        & \makecell[c]{cardiovascular disease\\glandular disease\\nervous disorder\\communicable disease\\inflammatory disease\\pycnosis\\skin disease\\cancer}
        & \makecell[c]{DGD, DCD, DSD\\GDG, GCG, GSG\\CDC, CGC, CSC\\SDS, SGS, SCS} \\
		\bottomrule
    \end{tabular}
    }
    \label{tab:dataset_statistics}
\end{table*}

\section{Baseline Descriptions}
\label{app:baselines}
Detailed descriptions of the baselines are as follows:
\begin{itemize}
\item  \textbf{HAN} \cite{jinyu:5} develops a hierarchical attention structure that includes node-level attention and semantic-level attention to fully consider the importance of node and metapath. However, the end-to-end training manner limits its ability to support generalizable heterogeneous graph learning.

\item  \textbf{HeCo} \cite{jinyu:45} employs cross-view contrastive mechanism, where the two views of heterogeneous graph, including network schema and metapath views, collaboratively supervise each other and finally learn high-level node
embeddings. Although it performs well in few-shot scenarios, its reliance on pretraining and fine-tuning for a specific heterogeneous graph significantly restricts its ability to generalize across different domains and tasks.

\item \textbf{WalkLM} \cite{jinyu:19} proposes leveraging random walk-based textual sequences to capture both complex attribute information and flexible topological structures of graphs in an unsupervised manner. However, it overlooks the rich semantic information in HGs.

\item  \textbf{GraphGPT} \cite{jinyu:11} proposes a dual-stage instruction tuning paradigm that aligns LLMs with graph structural knowledge, enabling generalization across unseen graphs in node classification tasks. However, GraphGPT only considers homogeneous graphs, neglecting the rich structural and semantic information present in HGs.

\item  \textbf{HiGPT} \cite{jinyu:10} is a heterogeneous graph instruction-tuning model, which introduces heterogeneity-aware graph instructions to help the LLM comprehend the complex relations in HGs. However, HiGPT leverages HGNNs to extract structural information as HG tokens, leading to biases in LLMs' understanding. Moreover, the node-level pretraining task limits its multi-task generalization ability, which further leads to suboptimal performance.

% \textbf{OFA} \cite{jinyu:41} utilizes a single LLM to encode textual descriptions of nodes and edges from different domain graphs in a unified manner. It employs nodes-of-interest (NOI) subgraphs and NOI prompt nodes to standardize various graph tasks. OFA performs well in cross-domain and cross-task scenarios. However, it does not take heterogeneous graphs into account.

% \item  \textbf{LLaGA} \cite{jinyu:12} employs neighborhood detail template and hop-field overview template to transform graph structure into sequences, and then maps node embeddings to token embedding spaces using a tuned projector. Although it can be generalizable to unseen datasets or tasks, LLaGA overlooks the prevalent heterogeneous graphs in real-world scenarios, and the use of the projector introduces noise into the LLM's understanding of graphs.
\end{itemize}

\section{Implementation Details}
\label{app:implement details}
Our experiments are carried out on a Linux OS with a single NVIDIA A800 GPU (80GB memory). The MLM4HG implementation relies on libraries including PyTorch \cite{jinyu:21}, PyTorch Geometric \cite{jinyu:22}, DGL \cite{jinyu:23}, scikit-learn \cite{jinyu:24}, and Transformers \cite{jinyu:25}. The model is fine-tuned for 3 epochs with a batch size of 16. During fine-tuning, the maximum token sequence length for input data is set to 512. The optimizer used is AdamW \cite{jinyu:26}, with a learning rate of 1e-5 and a weight decay of 0.01.

\section{Scalability Analysis}
\label{app:time complexity}
We discuss the time complexity and scalability of MLM4HG. MLM4HG employs a masked language model as the backbone, with time complexity scaling linearly with the token count in the corpus. For larger-scale graphs, longer metapath-based sequences may increase token sequence length and impact computational efficiency. To mitigate this, we randomly sampled 3 instances per metapath per node for the YELP dataset, which contains millions of edges, while still achieving promising results (Tables \ref{tab:main cd-ct few-shot} and \ref{tab:main cd-ct zero-shot}). Additionally, we provide detailed statistics in Table \ref{tab:time complexity}, including the number of nodes and edges, total token count (for both NC and LP), fine-tuning time per graph and inference time per batch (40 examples) across four datasets, further demonstrating MLM4HG’s scalability.

\begin{table}[htbp]
    \centering
    % \scriptsize
    \caption{Scalability analysis.}

    \resizebox{\textwidth}{!}{
    \begin{tabular}{c|c|c|c|c|c|c}
    \toprule
    
    & \textbf{Node Num.} & \textbf{Edge Num.} & \textbf{Fine-Tuning} & \textbf{Inference} & \textbf{NC Token Count} & \textbf{LP Token Count}\\
    \midrule
    
    IMDB & 11,616 & 34,212 & 338.68s & 1.0521s & 934,956 & 1,810,104\\
    DBLP & 26,128 & 239,566 & 381.69s & 1.3190s & 1,684,221 & 2,458,781\\
    PubMed & 63,110 & 246,518 & 452.92s & 1.2754s & 79,015 & 1,291,117\\
    YELP & 82,465 & 5,338,522 & 334.44s & 1.5963s & 1,364,293 & 1,794,996\\
    
    \bottomrule
    \end{tabular}
    }
    \label{tab:time complexity}
\end{table}

\section{Impact of Different LMs and LLMs}
\label{app:Impact of LMs}
The impact of different LMs is worth investigating. Considering the recent advances of causal language modeling (CLM) approaches\cite{jinyu:28, jinyu:27}, we not only evaluate MLM-based models, such as RoBERTa \cite{jinyu:18} and DistilRoBERTa \cite{jinyu:17}, but also explore the performance of LLaMA3 \cite{jinyu:27} (specifically, LLaMA-3.1-8B-Instruct) and Qwen3 \cite{jinyu:56} (specifically, Qwen3-8B) in our experiments. To ensure a fair comparison, we leverage instructions to restrict CLM-based models' predictions to the constrained target vocabulary. We also report the training time per batch for each model, with a batch size of 16. For MLM-based language models, full parameter fine-tuning are applied, while CLM-based models are fine-tuned using LoRA \cite{jinyu:49}.

\begin{table}[b]
    \centering
    % \scriptsize
    \caption{Average Micro-F1 scores for node classification and link prediction on IMDB and YELP with different LMs and LLMs. The best results are highlighted in bold and the second-best are underlined.}

    % \resizebox{0.5\textwidth}{!}{
    \begin{tabular}{c|c|c|c|c|c}
    \toprule
    
    \multicolumn{2}{c|}{\textbf{Task}} & \multicolumn{2}{c|}{\textbf{Node Classification}} & \multicolumn{2}{c}{\textbf{Link Prediction}}\\
    \midrule

    \textbf{LM} & \textbf{Time} & \textbf{IMDB} & \textbf{YELP} & \textbf{IMDB} & \textbf{YELP} \\
    \midrule
    
    DistilRoBERTa & 0.47s & \textbf{0.4622} & \underline{0.1772} & \textbf{0.9054} & \textbf{0.8378}\\
    RoBERTa & 0.68s & \underline{0.4417} & \textbf{0.2268} & \underline{0.8785} & \underline{0.7713}\\
    \midrule
    
    LLaMA3 & 2.28s & 0.3364 & 0.1693 & 0.8584 & 0.6903 \\
    Qwen3 & 2.02s & 0.2644 & 0.1475 & 0.6339 & 0.6013 \\
    \bottomrule
    \end{tabular}
    % }
    \label{tab:LMs and LM traing strageties}
\end{table}

We evaluated the impact of different LMs on the experimental results in node classification and link prediction tasks on the IMDB and YELP datasets. The experimental results are shown in Table \ref{tab:LMs and LM traing strageties}. Language models trained with MLM outperform those using CLM and require less training time. While CLM excels in many generative tasks \cite{jinyu:29}, MLM is better suited for our classical graph tasks, which are unified into a cloze-style format and thus demand a stronger focus on both left-to-right and right-to-left contextual information. In addition, as a lightweight variant of RoBERTa, DistilRoBERTa achieves competitive results while significantly reducing both the parameter count and training time. Therefore, we adopt DistilRoBERTa as the backbone model in our experiments.

\section{Hyperparameter Study}
\label{app:hyper_para}
\begin{figure}[htbp]
    \centering
    \includegraphics[width=1\linewidth]{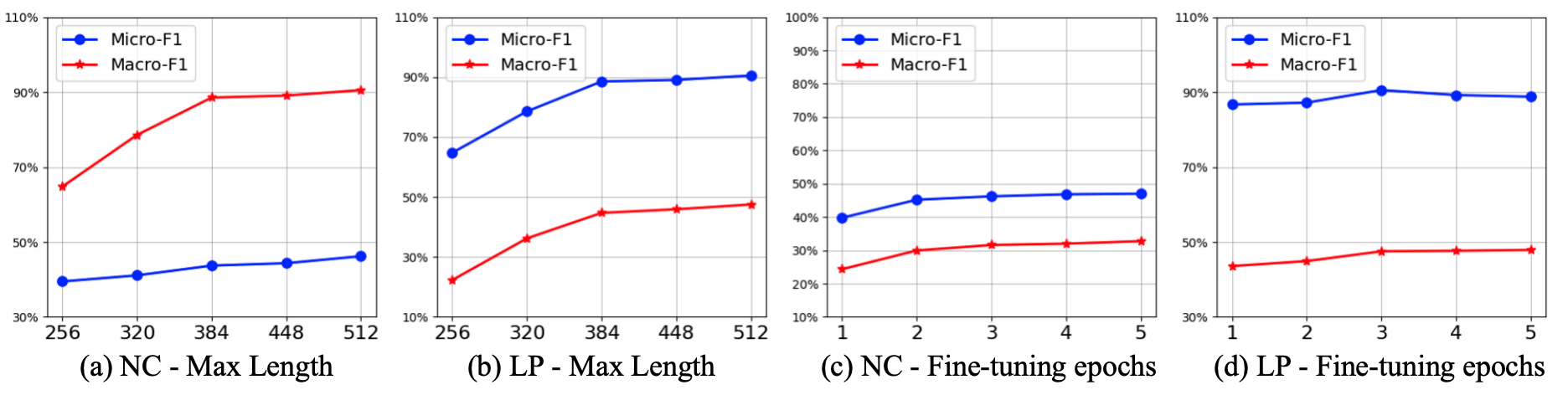}
    \caption{Average Micro-F1 and Macro-F1 scores for node classification (NC) and link prediction (LP) with different hyperparameter settings on IMDB dataset in the zero-shot scenarios.}
    \label{fig:hyper_para}
\end{figure}

% We evaluated the impact of two hyperparameters: maximum token sequence length (Max Length) and fine-tuning epochs on MLM4HG performance. As shown in Figure \ref{fig:hyper_para}, increasing the token sequence length enhances both node classification and link prediction, indicating that longer metapath-based sequences help LM capture more structural and semantic information. Additionally, as fine-tuning epochs increase, both node classification and link prediction tasks exhibit initial improvement followed by stabilization. Thus, we set the fine-tuning epochs to 3 in our experiments.
We conducted a detailed evaluation of two key hyperparameters: maximum token sequence length (Max Length) and the number of fine-tuning epochs (Fine-tuning epochs), to understand their influence on the performance of MLM4HG. As illustrated in Figure~\ref{fig:hyper_para}, extending the token sequence length consistently improves results for both node classification and link prediction tasks. This suggests that longer metapath-based sequences enable the language model to capture richer structural and semantic context from the heterogeneous graph. Furthermore, increasing the number of fine-tuning epochs initially leads to noticeable performance gains for both tasks, after which the results tend to stabilize. Based on this observation, we fixed the number of fine-tuning epochs to 3 in all reported experiments to balance performance and training efficiency.

\section{Zero-shot Link Prediction Evaluation Using AUC and AP}
\label{auc and ap for lp}
Table \ref{tab:auc and ap} presents the evaluation results of AUC and AP metrics, which serve as complementary measures for the zero-shot link prediction task. These results consistently exhibit the same trends as the F1 scores in Table \ref{tab:main cd-ct zero-shot}, further validating our conclusions.

\begin{table}[h]
    \centering
    % \scriptsize
    \caption{Average AUC and AP scores for link prediction in zero-shot scenarios. The best results are highlighted in bold and the second-best are underlined.}
    \resizebox{\textwidth}{!}{
    \setlength{\tabcolsep}{0.9mm}
    \begin{tabular}{c|c|c|c|c|c|c||c|c|c|c|c|c}
    \toprule
    
    {\textbf{Metric}} & \multicolumn{6}{c||}{\textbf{AUC}} & \multicolumn{6}{c}{\textbf{AP}} \\
    \midrule
    
    \textbf{Target} & \textbf{HAN} & \textbf{HeCo} & \textbf{WalkLM} & \textbf{GraphGPT} & \textbf{HiGPT} & \textbf{MLM4HG} & \textbf{HAN} & \textbf{HeCo} & \textbf{WalkLM} & \textbf{GraphGPT} & \textbf{HiGPT} & \textbf{MLM4HG} \\
    \midrule
    
    \textbf{IMDB} & 0.5194& 0.5047& 0.5438& 0.5836& \underline{0.6416}& \textbf{0.9471}& 0.1892& 0.1724& 0.2067& 0.2236& \underline{0.2822}& \textbf{0.7812}\\ 
    
    \textbf{DBLP} & 0.5423& 0.5302& 0.5634& 0.5942& \underline{0.6208}& \textbf{0.8160}& 0.1475& 0.1325& 0.1555& \underline{0.2292}& 0.2255& \textbf{0.4762}\\
    
    \textbf{PubMed} & 0.5167& 0.5106& 0.5242& 0.5366& \underline{0.5452}& \textbf{0.6886}& 0.1506& 0.1397& 0.1657& \underline{0.1997}& 0.1858& \textbf{0.3023}\\
    
    \textbf{YELP} & 0.5143& 0.5079& 0.5237& 0.5924& \underline{0.6967}& \textbf{0.8596}& 0.1296& 0.1236& 0.1451& 0.2084& \underline{0.3630}& \textbf{0.6254}\\
    \bottomrule
    
    \end{tabular}
    }
    \label{tab:auc and ap}
\end{table}

\section{Limitations and Broader Impacts}
\label{app:limitation}
We discuss the limitations and broader impacts of MLM4HG in this section. Although our exploratory experiments on metapath-based sequences in Section~\ref{sec:explore} demonstrate that MLM4HG is able to extract and leverage structural signals from such sequences for downstream tasks, the semantic information encoded in textual content remains crucial for language models. Therefore, applying LMs to non-textual graph data still requires further exploration. Moreover, additional investigation is needed to assess the model’s generalizability across a wider range of graph tasks.

MLM4HG is a simple yet effective framework for generalizable heterogeneous graph learning. Its potential impact extends to a variety of real-world applications involving HGs, such as social network analysis, knowledge representation, and recommender systems. While the ability to generalize across domains and tasks can enhance the development of robust AI systems, it may also introduce concerns. For example, the model could be misused in illegal activities, leading to unintended consequences such as the perpetuation of biases or the misuse of predictive insights. We encourage future work to focus on enhancing the transparency and fairness of graph language models, especially in sensitive applications.

\section{Case Study}
\label{app:case study}
In this section, we first provide some entries of the heterogeneous graph-based corpus used in our experiments. We then perform a visual analysis of the attention scores computed by the last layer of the model. Higher scores correspond to darker background colors, indicating that the relevant words have a greater impact on the prediction of \texttt{<mask>}.

\begin{figure}[t]
    \centering
    \includegraphics[width=1\linewidth]{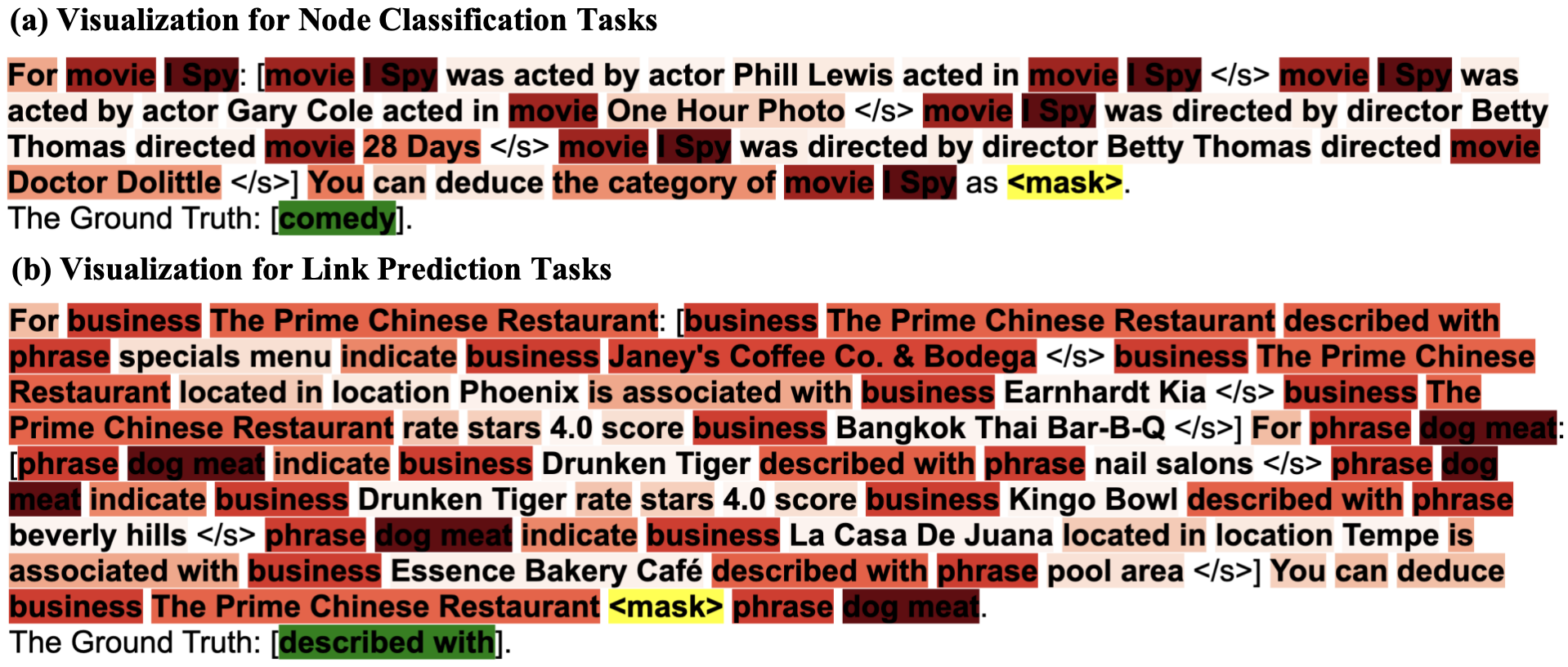}
    \caption{Visualization of the attention scores computed by the last layer of our MLM4HG for different graph tasks. Higher scores correspond to darker background colors, indicating that the relevant words have a greater impact on the prediction of the \texttt{<mask>} token.}
    \label{fig:case study}
\end{figure}

The entries for node classification and link prediction tasks are presented in Table \ref{tab:entries}. Fine-tuning the masked LM on such entries enables generalization across domains and graph tasks. Furthermore, we present a visualization of the attention scores computed by MLM4HG on the IMDB dataset. (1) Figure \ref{fig:case study} (a) shows the visualization results of node classification task for the movie \texttt{I Spy}. From the intensity of the colors, we observe that the model first focuses on the node's attribute information (name, \texttt{I Spy}) and node type (\texttt{movie}). Additionally, the model attends to information from neighboring nodes (\texttt{28 Days} and \texttt{Doctor Dolittle}), which belong to the same category as the target node. Finally, the model identifies the task as a node classification task (\texttt{the category of}), assisting in making the correct prediction. (2) Figure \ref{fig:case study} (b) illustrates the link prediction task between the source node business \texttt{The Prime Chinese Restaurant} and the destination node phrase \texttt{dog meat}. In addition to the attributes of the source and destination nodes, the model places more emphasis on the structural information within the metapath sequences, such as \texttt{<business described with phrase>}, \texttt{<phrase indicates business>}, and \texttt{<location is associated with business>}, ultimately providing an accurate prediction.

\begin{table}
    \centering
    \caption{Entries of the heterogeneous graph corpus for node classification (NC) and link prediction (LP) tasks.}
    \renewcommand{\arraystretch}{1.2}
    \resizebox{\textwidth}{!}{
    \begin{tabular}{c|c|p{0.9\textwidth}}
    \toprule
    Task & Dataset & Entries \\
    \midrule
    \multirow{4}{*}{\centering \textbf{NC}} & \textbf{IMDB} & \textbf{For movie I Spy}: [movie I Spy was acted by actor Phill Lewis acted in movie I Spy \texttt{</s>} movie I Spy was acted by actor Gary Cole acted in movie One Hour Photo \texttt{</s>} movie I Spy was directed by director Betty Thomas directed movie 28 Days \texttt{</s>} movie I Spy was directed by director Betty Thomas directed movie Doctor Dolittle \texttt{</s>}] You can deduce the category of movie I Spy as \texttt{<comedy>}. \\
    
    & \textbf{DBLP} & \textbf{For author Keith L. Doty}: [author Keith L. Doty write paper MICRONET: A Microcomputer Network System for Managing Distributed Relational Databases. was published in term managing publish paper Managing Healthcare Data Hippocratically. was written by author Ameet Kini \texttt{</s>} author Keith L. Doty write paper Extending Object-Oriented Concepts to Support Engineering Applications. was written by author Keith L. Doty \texttt{</s>}] You can deduce the category of author Keith L. Doty as \texttt{<0>}. \\

    & \textbf{PubMed} & \textbf{For disease dermatoses}: [disease dermatoses contain chemical trendline in disease tyrosination \texttt{</s>} disease dermatoses contain chemical F16357 in disease MDCP \texttt{</s>} disease dermatoses is caused by gene MicroRNA-146a causing disease dermatoses \texttt{</s>} disease dermatoses is caused by gene DmPpt1 causing disease autoimmune cytopenias \texttt{</s>}] You can deduce the category of disease dermatoses as \texttt{<skin disease>}. \\

    & \textbf{YELP} & \textbf{For business The Living Room}: [business The Living Room described with phrase 15k miles indicate business Hollywood Beauty Eyebrow Threading \texttt{</s>} business The Living Room described with phrase mexican street corn indicate business Burger Theory \texttt{</s>} business The Living Room located in location Scottsdale is associated with business Daphne's California Greek \texttt{</s>} business The Living Room rate stars 4.0 score business Bell Lexus North Scottsdale \texttt{</s>}] You can deduce the category of business The Living Room as \texttt{<sandwiches>}.\\
    \midrule

    \multirow{4}{*}{\centering \textbf{LP}} & \textbf{IMDB} & \textbf{For movie Pixels}: [movie Pixels was acted by actor Josh Gad acted in movie The Angry Birds Movie \texttt{</s>} movie Pixels was acted by actor Josh Gad acted in movie Ice Age: Continental Drift \texttt{</s>} movie Pixels was directed by director Chris Columbus directed movie Bicentennial Man \texttt{</s>}] \textbf{For actor Adam Sandler}: [actor Adam Sandler acted in movie Grown Ups was directed by director Dennis Dugan directed movie You Don't Mess with the Zohan was acted by actor Kevin Nealon \texttt{</s>} actor Adam Sandler acted in movie Hotel Transylvania was acted by actor Steve Buscemi \texttt{</s>}] You can deduce movie Pixels \texttt{<was acted by>} actor Adam Sandler. \\
    
    & \textbf{DBLP} & \textbf{For term relational}: [term relational publish paper Integrating Pattern Mining in Relational Databases. was written by author Toon Calders write paper Mining rank-correlated sets of numerical attributes. was published in term correlated \texttt{</s>} term relational publish paper Updating Relational Databases through Object-Based Views. was published in term updating \texttt{</s>}] \textbf{For paper Learning Probabilistic Relational Models.}: [paper Learning Probabilistic Relational Models. was published in term relational publish paper Using relational knowledge discovery to prevent securities fraud. \texttt{</s>} paper Learning Probabilistic Relational Models. was received by conf IJCAI receive paper Parametric Kernels for Sequence Data Analysis. \texttt{</s>}] You can deduce term relational \texttt{<publish>} paper Learning Probabilistic Relational Models. . \\ 

    & \textbf{PubMed} & \textbf{For gene U2af1-rs1}: [gene U2af1-rs1 \texttt{</s>}] \textbf{For disease neuroendocrine tumours}: [disease neuroendocrine tumours contain chemical ammonium phosphate in disease neuroendocrine tumours \texttt{</s>} disease neuroendocrine tumours contain chemical DiMeIQx in disease chronic hypertension \texttt{</s>} disease neuroendocrine tumours \texttt{</s>} disease neuroendocrine tumours is caused by gene Cathepsin A causing disease Bone Marrow Failure \texttt{</s>} disease neuroendocrine tumours is caused by gene S2242 causing disease nasopharyngeal secretions \texttt{</s>}] You can deduce gene U2af1-rs1 \texttt{<causing>} disease neuroendocrine tumours. \\

    & \textbf{YELP} & F\textbf{or phrase great taste}: [phrase great taste indicate business Mika's Greek rate stars 4.0 score business Thai Basil described with phrase shrimp risotto \texttt{</s>} phrase great taste indicate business Sonic Drive-In rate stars 5.0 score business Rubio's described with phrase joe clark \texttt{</s>} phrase great taste indicate business Fired Pie described with phrase mind's eye \texttt{</s>} phrase great taste indicate business Dj's Bagel Cafe described with phrase free appetizer \texttt{</s>}] \textbf{For business Sal's Tuscan Grill}: [business Sal's Tuscan Grill rate stars 4.0 score business Bosa Donuts \texttt{</s>} business Sal's Tuscan Grill rate stars 4.0 score business uBreakiFix \texttt{</s>} business Sal's Tuscan Grill described with phrase iced americano indicate business El Palacio \texttt{</s>} business Sal's Tuscan Grill described with phrase lamb vindaloo indicate business Chapman Volkswagen Scottsdale \texttt{</s>}] You can deduce phrase great taste \texttt{<indicate>} business Sal's Tuscan Grill. \\
    \bottomrule
    
    \end{tabular}
    }
    \label{tab:entries}
\end{table}

\section{Examples and Parsing Results}
\label{app:examples and parsing results}
\begin{figure}[htbp]
    \centering
    \includegraphics[width=1\linewidth]{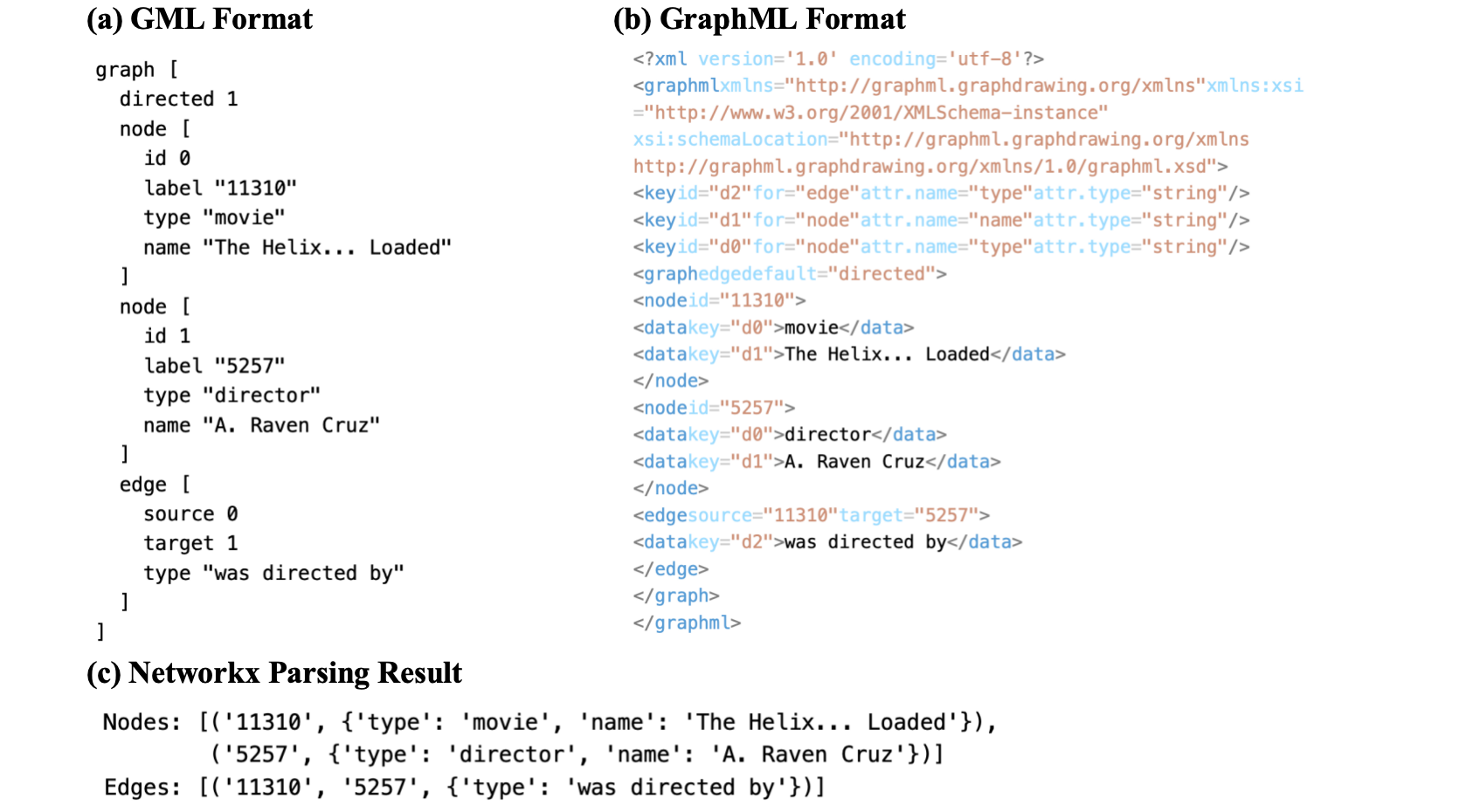}
    \caption{Examples and parsing results of GML and GraphML formats.}
    \label{fig:gml}
\end{figure}

Examples and parsing results of GML and GraphML formats are provided in Figure \ref{fig:gml}.

%%%%%%%%%%%%%%%%%%%%%%%%%%%%%%%%%%%%%%%%%%%%%%%%%%%%%%%%%%%%

\end{document}